\documentclass[12pt]{article}

\usepackage{color}
\usepackage{latexsym}
\usepackage{epsfig,amssymb,euscript}
\usepackage{amsmath,slashed,cancel}
\usepackage{array,calc,epsfig}
\usepackage[linktocpage]{hyperref}
\usepackage{pstricks}
\usepackage{bbm}

\def\be{\begin{equation}}
\def\ee{\end{equation}}
\def\bseq{\begin{subequations}}
\def\eseq{\end{subequations}}

\def\bea{\begin{eqnarray}}
\def\eea{\end{eqnarray}}
\newcommand\bbone{\ensuremath{\mathbbm{1}}}

\numberwithin{equation}{section} 

\def\d         {{\rm d}}

\def\cala         {{\cal A}}

\def\cald         {{\cal D}}

\def\calf         {{\cal F}}

\def\calj         {{\cal J}}

\def\call         {{\cal L}}
\def\calm         {{\cal M}}
\def\caln         {{\cal N}}
\def\calo         {{\cal O}}

\def\calt         {{\cal T}}

\def\calw         {{\cal W}}
\def\calz         {{\cal Z}}

\def\itzero         {{\it 0}}
\def\itone         {{\it 1}}

\def\del          {\partial}
\def\delbar       {\bar\partial}

\def\tr           {\mathop{\rm Tr}}
\def\Re           {{\rm Re\hskip0.1em}}
\def\Im           {{\rm Im\hskip0.1em}}

\def\sqr#1#2{{\vcenter{\vbox{\hrule height.#2pt
 \hbox{\vrule width.#2pt height#1pt \kern#1pt \vrule width.#2pt}\hrule
 height.#2pt}}}}

\newcommand{\hlt}[1]{\textcolor{blue}{#1}}

\def\d{\text{d}}

\def\slashchar#1{\setbox0=\hbox{$#1$}           
\dimen0=\wd0                                 
\setbox1=\hbox{/} \dimen1=\wd1               
\ifdim\dimen0>\dimen1                        
\rlap{\hbox to \dimen0{\hfil/\hfil}}      
#1                                        
\else                                        
\rlap{\hbox to \dimen1{\hfil$#1$\hfil}}   
/                                         
\fi}

\begin{document}
\font\cmss=cmss10 \font\cmsss=cmss10 at 7pt

\vskip -0.5cm
\rightline{\small{\tt ITEP-TH-56/10}}
\rightline{\small{\tt ROM2F/2010/21}}

\vskip .7 cm

\hfill
\vspace{18pt}
\begin{center}
{\Large \textbf{D-brane non-perturbative effects  \\ and geometric deformations }}
\end{center}

\vspace{6pt}
\begin{center}
{\large\textsl{Anatoly Dymarsky $^a$ and Luca Martucci $^{b,c}$}}

\vspace{25pt}
\textit{\small $^a$ School of Natural Sciences, Institute for Advanced Study,\\
                   Princeton, NJ, 08540}\\ \vspace{6pt}
\textit{\small  $^b$ I.N.F.N. Sezione di Roma ``TorVergata'' \&\\  Dipartimento di Fisica, Universit\`a di Roma ``TorVergata", \\
Via della Ricerca ScientiÞca, 00133 Roma, Italy }\\  \vspace{6pt}
\textit{\small  $^c$ Arnold Sommerfeld Center for Theoretical Physics,\\ LMU M\"unchen,
Theresienstra\ss e 37, D-80333 M\"unchen, Germany}\\  \vspace{6pt}
\end{center}

\vspace{12pt}

\begin{center}
\textbf{Abstract}
\end{center}
We study how non-perturbative dynamics on D-branes affects the ten-dimensional geometry. We show that a gaugino condensate  changes the complex and the symplectic structures of the original manifold by deforming the supersymmetry conditions.
The cases of D5, D6 and D7-branes are discussed in detail. In the latter case we find the explicit form of the resulting back-reacted background at linear order in the gaugino condensate.
\vspace{4pt} {\small

\noindent }

\vspace{1cm}

\thispagestyle{empty}

\vfill
\vskip 5.mm
\hrule width 5.cm
\vskip 2.mm
{\small
\noindent e-mail: dymarsky@ias.edu\ , luca.martucci@roma2.infn.it
}

\newpage

\setcounter{footnote}{0}

\tableofcontents

\newpage

\section{Introduction}

Non-perturbative effects play an important role in many branches of string theory. In the case of the phenomenology-motivated  settings, the non-perturbative effects are usually approached from the low-energy four-dimensional perspective: the internal space is typically compactified and at low energy the theory reduces  to some four-dimensional effective theory. The latter usually involves the moduli of the internal space plus some gauge fields.

However, the non-perturbative phenomena are expected to have an effect already at the level of the ten-dimensional geometry.
Take for example a string theory compactification which includes a confining gauge sector, with a confining scale $\Lambda_{\rm conf}$.  For large $\Lambda_{\rm conf} \gtrsim \Lambda_{\rm KK}$, one expects that a proper description of the non-perturbative strongly coupled dynamics  should necessarily involve the complete ten-dimensional theory. When $\Lambda_{\rm conf}$ is reduced to $\Lambda_{\rm conf}\ll \Lambda_{\rm KK}$,  at low energies the ten-dimensional description should eventually match the effective four-dimensional one. In particular, in the ten-dimensional supergravity regime when the stringy corrections can be neglected one expects a direct modification of vacuum equations determining the geometrical structure and the matter content of the internal space.

A proper understanding of this problem can become important in the so-called local models which use the `bottom-up' approach focusing on a local patch  where some interesting physics takes place. The issues related to the understanding of the complete compactification are usually postponed. For instance, in the D3-brane inflationary models of \cite{KKLMMT} a deformation of the original classical warped conifold background \cite{KW,ks} is introduced in order to get a more realistic inflation potential \cite{Baumann:2007ah}.\footnote{Another example when such a deformation can be important is the engineering of the realistic Yukawa couplings on the intersecting seven-branes \cite{cecotti09,npYuk}.}
These models are based on the no-scale vacua of \cite{gkp} which do not a priori allow for such deformations at the classical level. Hence, from the point of view of the complete compact internal space,  deformations of this kind cannot be considered as perturbative ones. Rather they affect physics at the KK-scale, drastically changing the original background.  It is then interesting to understand how the deformations of the classical geometry can be {\em dynamically} generated by non-perturbative quantum effects.

In  type II settings with  D-brane non-perturbative effects, the above problem has been already addressed in \cite{tentofour,bau10} following two different approaches. One of the goals of this paper is to revisit the considerations of \cite{tentofour,bau10} combining them into a unifying picture. For the sake of clarity and concreteness, we will focus on several specific cases: the non-perturbative effects generated by gaugino
condensation on D5, D6 and D7 branes wrapping internal two-, three- and four-cycles respectively.

The backreaction of D5-branes wrapping a two-cycle in a non-compact CY-space  has been previously considered in the literature in the context of the geometric transition and the gauge/gravity correspondence \cite{ks,mn,vafa,vafa2,martellimalda}. In the prototypical setting, $N$ D5-branes are wrapping the rigid two-cycle at the tip of the resolved conifold and are described by an effective four-dimensional SYM theory which undergoes gaugino condensation. In the large $N$-limit, such a theory is described by a dual background in which the D-branes have disappeared and the gaugino condensation is represented by a deformation of the complex structure of the background. Hence, the background undergoes a geometric transition from the resolved to deformed conifold (see \cite{martellimalda} for a recent discussion). One can interpret this as a `backreaction' of the non-perturbative effect and in this paper we focus on this point of view. By following the general formalism introduced in \cite{tentofour}, we will provide an explanation of the direct dynamical origin of such an effect within the local ten-dimensional supergravity approach without relying on any compactification effects or using  any holographic argument.  In turn, the well-understood holographic viewpoint will provide a check of  our approach. As it is rooted in the supegravity description, our approach has an advantage of being  applicable to a more general set of situations going beyond the conifold geometry.

Furthermore, we will  also see how the very same approach works in mirror symmetric case with $N$ D6-branes wrapping rigid three-cycles. The simplest example of such a situation is when $N$ D6-branes wrap the three-sphere at the tip of the deformed conifold. It is the original example considered in \cite{vafa}. The results reached there are in perfect agreement with what we will obtain using our approach.

Reassured by the insight gained  from the cases involving D5 and D6-branes,  we will move on to study the effects of  gaugino condensation on D7-branes.  Similarly to the case with  D5 and D6-branes, we describe how the  non-perturbative effect of the D7-branes can be encoded as a backreaction of the underlying geometry.  In particular, it appears that at the leading order the  non-perturbative physics on the D7-branes has two effects: the deformation of the bulk integrable complex structure into a {\em generalized complex} structure \cite{tentofour} and the  generation of  IASD three-form flux \cite{bau10}.
We will provide a unifying framework that includes both points of view. Moreover, this will allow us to improve our understanding of the geometrization, proposed in \cite{tentofour} and \cite{bau10}, of the non-trivial superpotential for the mobile D3-branes  induced by  the D7-branes \cite{ganor96,baumann0607} supporting a gaugino condensate.


\section{Geometric deformations from  condensing D5-branes}
\label{sec:D5gen}

In this section we focus on compactifications with $N$ D5-branes wrapping an internal rigid two-cycle. We will first review the conditions that the internal six-dimensional geometry obeys at the classical level. Then we explain how these conditions can be obtained from a
superpotential which depends on the KK modes associated with the compactification \cite{tentofour}. We will then incorporate the effect of gaugino condensation and discuss some of its physical implications.

\subsection{The classical background}
\label{sec:D5}

Here we consider the supergravity backgrounds with the ten-dimensional space $X_{10}=X_4\times M$, where $X_4$ is the four-dimensional Minkowski space and $M$ is the internal, compact or non-compact, manifold. In the internal space we allow for possible D5-branes. In the case of compact $M$, O5-planes would then be necessary to cancel the net charge, and $M$ would be the associated covering space.

In this setting, four-dimensional $\caln=1$ supersymmetry implies that the background can be characterized as an SU(3)-structure manifold, as follows.\footnote{These vacua are called type C in the classification of \cite{GranaReview}.}  The string-frame metric can be written in the form
\be\label{D5metric}
\d s^2=e^{\phi}\d x^\mu\d x_\mu+\d s^2_M
\ee
where $\phi$ is the dilaton, not necessary constant.  The internal space is complex and has a holomorphic $(3,0)$-form $\Omega$
\be\label{holsusy}
\d\Omega=0\ .
\ee
On the other hand the fundamental two-form $J_{mn}:=g_{mk}I^k{}_n$ (where $I^m{}_n$ is the complex structure defined by $\Omega$) is not closed, but rather it must obey the following conditions\footnote{\label{foot:normomega} The normalization of $\Omega$ and $J$ is given by
$\text{vol}_M=(1/3!)J\wedge J\wedge J=-(i/8)\, e^{-\phi}\Omega\wedge \bar\Omega$. An alternative, more physical way to fix the normalization of $\Omega$ is that $\Omega$ calibrates D5-brane domain wall, i.e.\ a BPS domain wall obtained by wrapping a D5-brane on a three-cycle.}
\be\label{susycondD5}
\d(e^\phi J)=-e^{2\phi}*F_{\it 3}\ , \qquad \d(J\wedge J)=0\ .
\ee
Notice that the first relation in (\ref{susycondD5}) defines $F_{\it 3}$ in terms of the SU(3)-structure data and the dilaton. In addition, the flux $F_{\it 3}$ should obey the Bianchi Identity (BI)
\be\label{D5delta}
\d F_{\it 3}= \,\ell_s^2\,\Big[\sum_{a\in \text{D5-branes}}\delta^{{\it 4}}_{D_a} - \sum_{b\in \text{O5-planes}}\delta^{\it 4}_{O_b} \Big]
\ee
where $\ell_s=2\pi\sqrt{\alpha^\prime}$ and $D_a$ and $O_b$ are the holomorphic two-cycles wrapped by the D5-branes and O5-planes respectively. In (\ref{D5delta}),  $\delta^{{\it 4}}_{D}$ refers to the  internal delta-like  form   localized on the two-cycle $D$.\footnote{\label{fn:delta} A formal definition of the delta-like $k$-form $\delta^{{\it k}}_\Sigma$  on $M$, associated to a $(6-k)$-dimensional surface $\Sigma$, is as follows:  for any $(6-k)$-form $\omega$ on $M$, $\int_M\omega\wedge \delta^{\it k}_\Sigma\equiv \int_\Sigma\omega$.}

A crucial point is that  the ten-dimensional supersymmetry conditions, including (\ref{holsusy}),  can be obtained as F-flatness and D-flatness conditions derived from a superpotential and a K\"ahler potential, which can be seen as functionals of the complete tower of the KK-modes associated with  the compactification. This has been demonstrated  for the most general supersymmetric background with Minkowski or AdS four-dimensional space in \cite{tentofour} using the formalism of generalized complex geometry \cite{hitchin,gualtieri}. Here we do not need all that machinery, as we will focus just on the supersymmetry condition (\ref{holsusy}). This can be obtained from the
superpotential \cite{grana1,tentofour}\footnote{Let us emphasize  that (\ref{truncsupD5}) should be regarded as a superpotential of the superconformal formulation of the four-dimensional supergravity. In particular $\calw$  has fixed normalization, while in Einstein supergravity the superpotential $\calw_{\rm E}$ is defined up to a K\"ahler transformation. The two are related by $\calw=e^{K/2}\calw_{\rm E}$. See \cite{tentofour,effsugra} for more details.}
\be\label{truncsupD5}
\calw=-\frac{\pi}{\ell_s^8}\int_M \Omega\wedge \big(F_{\it 3}+ie^{-\phi}\d J\big)\ .
\ee
The overall minus sign is physically irrelevant  as it can be reabsorbed into a redefinition of $\Omega$, and has been chosen for notational consistency with the following sections.

We consider (\ref{truncsupD5}) (or, better yet, its generalization given in \cite{tentofour}, see (\ref{gensup}) below) as a `microscopic' superpotential, which includes information on the internal space, taking into account all KK modes without restricting to the low-energy light fields. In particular, for what concerns us, it is sufficient to recognize that $\calw$ depends holomorphically on the
`chiral field'
\be
\calt_{\it 2}:=e^{-\phi}J-iC_{\it 2}\ .
\ee
Clearly, $\calt_2$ is not a standard low-energy chiral field but rather an internal two-form which can be thought of as encoding an entire  KK-tower of chiral fields.

Considering a small fluctuation $\delta\calt_{\it 2}$, we get the following corresponding variation of the superpotential
\be
\delta \calw=-\frac{i\pi}{\ell_s^8}\int_M \d\Omega\wedge \delta\calt_{\it 2}\ .
\ee
Assuming a compactification with flat four-dimensional space, we obtain the following F-term  associated to $\calt_{\it 2}$
\be\label{Ftree}
\calf_{\calt_{\it 2}}:= \frac{\delta \calw}{\delta\calt_{\rm 2}}\,= -\,\frac{i \pi }{\ell_s^8}\,\d\Omega\ .
\ee
Here and in the following we restrict for simplicity to the case of the Minkowski $X_4$. Then supersymmetry requires  $\calw=0$ and the F-term associated with the chiral field $\phi^i$  is simply $\calf_i=\partial\calw/\partial\phi^i$. We recover the same expression in the rigid limit, when $\calw$ may not vanish but the corresponding term is suppressed by the Planck mass.
Then,  the supersymmetry F-flatness condition
$\calf_{\calt_{\it 2}}=0$
indeed  reproduces (\ref{holsusy}). It is important to notice that the F-term $\calf_{\calt_{\it 2}}$ is crucially associated to the massive KK-modes encoded in $\calt_{\it 2}$.


\subsection{Gaugino condensation and supersymmetry}
\label{sec:D5susy}

We would like to see how the IR strongly coupled dynamics on a stack of $N$ D5-branes modify the bosonic ten-dimensional supersymmetry conditions. Our primary  focus will be on  the condition (\ref{holsusy}). In a straightforward approach we should investigate how the complete ten-dimensional supersymmetry transformations are modified in the presence of the non-vanishing expectation value of the D5-brane gaugino bilinear, and how the resulting modified Killing spinor conditions translate into a corresponding modification of  the classical supersymmetry conditions reviewed in section \ref{sec:D5}.  However, this approach presents technical difficulties related to our ignorance of the coupled bulk and D-branes supersymmetry transformations. We will then follow an indirect derivation of the modified ten-dimensional conditions. Our approach will have the advantage of admitting a clear four-dimensional interpretation.

We already saw that (\ref{holsusy}) can be rephrased using the four-dimensional language as the vanishing of the F-term.
Now our strategy will be to compute how $\calf_{\calt_{\it 2}}$  is affected by the gaugino condensate within the formalism adopted in the previous section.
In general the backgrounds with compact internal space may develop an instability due to non-perturbative effect.  To avoid dealing with this issue we simply consider the limit of a non-compact internal space. This corresponds to having a rigid four-dimensional theory and will allow us to preserve supersymmetry.  The issues specific to the compactification will be discussed in  section \ref{sec:susyb}.

\bigskip

Let us first recall how things go in a standard four-dimensional $\caln=1$ theory. For simplicity, we will work with the rigid superspace formalism. We consider a theory with chiral superfields $\phi^i$ and  a gauge multiplet sector. Then the general supersymmetric effective Lagrangian contains the following chiral contributions\footnote{\label{foot:trnorm} The generators $T_a$ of the gauge group $G$ are defined such that  $\tr T_a T_b=\frac12 \delta_{ab}$. Hence, $\tr W^\alpha W_\alpha=\frac12 W^{a\alpha} W^a_\alpha$.}
\be
\int \d^2\theta\, \calw(\phi)+\frac{1}{8\pi}\int \d^2\theta\, \alpha(\phi)\tr W^\alpha W_\alpha \, +\, \text{c.c.}
\ee
where $\calw$ is the superpotential for the chiral fields, while $W_\alpha$ is the superfield whose lowest component is the gaugino:
\be
W_\alpha=-i\lambda_\alpha+\ldots
\ee
 Furthermore the lowest component of $\alpha(\phi)$ gives the SYM coupling and the theta-angle ($\alpha$ is related to the conventional coupling $\tau$  through $\tau\equiv i\alpha$)
\be\label{SYMcoupling}
\alpha(\phi)=\frac{4\pi}{g^2_{\rm YM}}-i\,\frac{\theta_{\rm YM}}{2\pi}
\ee
as can be seen from the standard superspace integration
\bea\label{SYMexp}
&&\frac{1}{8\pi}\int \d^2\theta\, \alpha(\phi)\tr W^\alpha W_\alpha +\,\text{c.c.}=\cr
&&-\frac{1}{2g^2_{\rm YM}}\,\tr F_{\mu\nu}F^{\mu\nu}+\frac{\theta_{\rm YM}}{32\pi^2}\epsilon^{\mu\nu\rho\sigma}\tr F_{\mu\nu}F_{\rho\sigma}-\frac{2i}{g^2_{\rm YM}}\,\tr\lambda\sigma^\mu\partial_\mu\bar\lambda +\ldots
\eea
We employ the standard two-components notation for the spinor indices as in \cite{bw}.
Of course, $g_{\rm YM}$ and $\theta_{\rm YM}$ must be considered as functions of the chiral fields $\phi^i$.
Now, the complete expressions for the  F-terms $\calf_{\phi^i}$ associated with $\phi^i$ will contain the fermionic bilinear $\lambda^\alpha\lambda_\alpha$
\be
\calf_{\phi^i}=\partial_i\calw-\frac{1}{8\pi}\,(\partial_i\alpha) \, \tr\lambda^\alpha\lambda_{\alpha}+\ldots
\ee
To describe the  gaugino condensate we introduce the superfield
\be\label{defS}
S=-\frac{1}{16\pi^2}\tr W^\alpha W_\alpha=\frac{1}{16\pi^2}\tr \lambda^\alpha \lambda_\alpha+\ldots
\ee
In presence of the non-vanishing gaugino condensate we have
\be\label{gcond}
\langle S\rangle =\frac{1}{16\pi^2}\tr \langle \lambda^\alpha \lambda_\alpha\rangle\neq 0
\ee and the F-flatness condition takes the form
\be\label{modFterms}
\calf_{\phi^i}=\partial_i\calw- 2\pi \, \langle S\rangle\,\partial_i\alpha=0\ .
\ee

In passing, let us recall that  the very same equations can be formally obtained from a different perspective (still in the rigid  limit $M_{\rm P}\rightarrow \infty$).  Take the gauge group $G=$SU($N$). Then we may start with the  superpotential
\be
\tilde\calw(\phi,S) =\calw(\phi)-2\pi \alpha(\phi)\, S+NS(1-\log \frac{S}{\mu_0^3})
\ee
where $\mu_0$ indicates  the scale at which the holomorphic coupling $\alpha$ is defined.
Once one identifies  $\Lambda=\mu_0\,e^{-\frac{2\pi\alpha}{3N}}$,  $\tilde\calw(\phi,S)$ is nothing but $\calw(\phi)$ corrected by the Veneziano-Yankielowicz  superpotential \cite{Veneziano:1982ah}, governing the low-energy dynamics of $S$. We could now pass to the effective superpotential obtained by integrating out $S$,
\be\label{effsupd5}
\tilde\calw_{\rm eff}(\phi) =\calw(\phi)+N\,\mu_0^3\, e^{-\frac{2\pi\alpha(\phi)}{N}}
\ee
This is the approach implicitly followed in \cite{tentofour} (see \cite{Lust:2005cu} for the early work based on a similar idea). However it   presents some conceptual subtleties related to the proper interpretation of $\tilde\calw_{\rm eff}$. In particular,    the extrapolation  of $\tilde\calw_{\rm eff}$ to a  ten-dimensional framework seems not to be a priori justified. For this reason, we prefer the logic followed above, in which we simply assume a non-trivial expectation value of the gaugino condensate without specifying its dynamical origin.  Notice also that this approach can be applied to the more general gauge groups and matter content on the D-branes.

\bigskip

Let us then go back to our ten-dimensional problem. Our task is to understand how the gaugino condensation  on a stack of $N$ D5-branes wrapping a rigid two-cycle $D$ in $M$ modifies the {\em ten-dimensional} supersymmetry equations reviewed in section \ref{sec:D5}. As already mentioned, attacking this problem directly appears difficult. We will circumvent these difficulties by using the re-interpretation of the ten-dimensional supersymmetry conditions in terms of the four-dimensional language. The question now is: how does the presence of the gaugino condensate on the D5-branes modify the F-flatness condition (\ref{Ftree})?

By dimensionally reducing the bosonic D5-brane DBI+CS action on  $D$ we obtain
\be\label{D5action}
S_{\rm D5}=-\frac{1}{8\pi\ell_s^2}\int_{D}e^{-\phi}J\int_{X_4} \d ^4 x\,\sqrt{-\det g_4} \,\tr F_{\mu\nu}F^{\mu\nu}+\frac{1}{4\pi\ell_s^2}\int_{D}C_{\it 2}\int_{X_4}\tr  F\wedge F +\ldots
\ee
where we have only indicated the massless 4D SU(N) gauge sector and `$\ldots$' stands for the terms containing other KK  degrees of freedom on the D5-branes with masses of order the inverse size of  $D$.  It is important to keep in mind that we are considering the D5 theory as a six-dimensional one that couples to the complete ten-dimensional bulk closed string sector. In particular, near a supersymmetric vacuum configuration, the D-brane sector should organize into a massless vector multiplet plus a tower of massive chiral and vector multiplets. Hence,  we are just reorganizing the higher dimensional theory as a four-dimensional theory of massless as well as massive KK-modes.

By comparing (\ref{D5action}) and (\ref{SYMexp}) one can easily identify the holomorphic gauge coupling associated with the massless SYM sector
\be\label{alphacalt}
\alpha\equiv \alpha(\calt)= \frac{1}{\ell_s^2}\int_{D}(e^{-\phi}J-i C_{\it 2})\equiv \frac{1}{\ell_s^2}\int_{D}\calt_{\it 2}\ .
\ee
It turns out to be depend only on $\calt_2$.
This is a bare coupling defined at the natural cut-off scale $\mu_0$ at which the ten-dimensional effective action breaks down. This scale can be roughly identified with the string scale $1/\ell_s$.

Now we need only adapt  (\ref{modFterms}) to our ten-dimensional setting, by using (\ref{truncsupD5}) and (\ref{alphacalt}).  It is easy to see that the F-term (\ref{Ftree}) gets modified to
\be\label{Fgaugino}
\calf_{\calt_{\it 2}}\,=\, \frac{\delta \calw}{\delta\calt_{\rm 2}}- 2\pi \, \langle S\rangle\,\frac{\delta \alpha}{\delta\calt_{\rm 2}}\,\equiv -\,\frac{i \pi }{\ell_s^8}\,\d\Omega-\frac{2\pi}{ \ell_s^{2}}\, \langle S\rangle \,\delta^{\it 4}_{D}
\ee
Hence the associated F-flatness condition $\calf_{\calt_{\it 2}}=0$ reads
\be\label{defomega_2}
\d\Omega={2i\,\ell^6_s}\, \langle S\rangle \,\delta^{\it 4}_{D}\ .
\ee
The  gaugino condensate has modified the supersymmetry condition (\ref{holsusy}) into (\ref{defomega_2}).

Let us mention here that (\ref{defomega_2}) is a subcase of a more general formula obtained in \cite{tentofour}.  However the procedure followed here allows for a sharper physical interpretation. In particular, this interpretation  suggests that the remaining supersymmetry conditions are left unchanged. This is because the holomorphic gauge coupling (\ref{alphacalt}) does not depend on other chiral fields whose F-terms could be associated to the remaining supersymmetry conditions.
Of course there could be subtleties affecting this naive conclusion. Indeed, we will see that the remaining equations could be modified by the localized  terms induced by gaugino condensation, which can be reabsorbed in a redefinition of the internal fluxes. This effect is known in the heterotic framework \cite{drsw,hor96} (see also \cite{Frey}) as discussed in the next section.  In the type II framework we will see a concrete example of this effect in the setting with D7-branes considered in sections \ref{sec:D7} and \ref{sec:firstord}.

In summary, our claim is the following. In the ten-dimensional supergravity approximation, a  non-vanishing expectation value for the four-dimensional gaugino bilinear associated with the  four-dimensional SYM sector coming from the  $N$ D5-branes wrapping a rigid two-cycle $D$, deforms  the {\em ten-dimensional} supersymmetry condition (\ref{holsusy})  into (\ref{defomega_2}). We also
 expect the  other conditions not to be modified or to be modified only by the localized terms. Let us stress again that we are considering the limit in which the internal space decompactifies and the theory is genuinely ten-dimensional or, in other words, we are just focusing on a local patch around the two-cycle wrapped by the condensing D5-branes. We will come back to the compactification effects  in section \ref{sec:susyb}.


\subsection{The analogy with the heterotic M-theory case}
\label{sec:hetM}

What we have done so far is conceptually very similar, although procedurally and technically different, to the approach followed by ${\rm Ho\check{r}ava}$ in \cite{hor96}. There the very same problem was studied in the context of heterotic M-theory \cite{HW}. In that case the internal space is a seven-dimensional space of the form $N=M\times (S^1/\mathbb{Z}_2)$, with $M$ six dimensional CY, and two $E_8$ SYM sectors localized at the orbifold fixed points, in the eleventh direction $y^{11}$.

As we did above, ${\rm Ho\check{r}ava}$ first employed the local point of view, in which $S^1/\mathbb{Z}_2 $ decompactifies to $\mathbb{R}/\mathbb{Z}_2$ with only one $E_8$ gauge sector localized at $y^{11}=0$.  In the heterotic M-theory case the complete low-energy eleven-dimensional supersymmetric theory coupled to the localized gauge sector is under control and one can directly compute the effects of the gaugino condensate on the Killing spinor equations. If $\eta$ and $\eta+\Delta\eta$ denote the internal Killing spinor before and after the gaugino condensate is turned on, the result of \cite{hor96} (see equation (3.6) therein) is that $\Delta\eta$ must satisfy a Killing condition of the form
\be\label{Mhetkilling}
\partial_{y}\Delta\eta\,\propto\, \langle S\rangle\, \delta(y)\ .
\ee
Comparing it with (\ref{defomega_2}) we see that the two relations share the same structure. The analogy is strengthened if we recall that in the D5-brane setting $\Omega$ is constructed  as a bilinear of the internal Killing spinor: $\Omega_{mnp}\simeq \eta^T\gamma_{mnp}\eta$. Hence, both (\ref{defomega_2}) and (\ref{Mhetkilling}) predict a delta-function-like contribution to the derivative of the Killing spinor at the location of the defect on which the strongly coupled gauge sector is localized.

Hence, our results are similar to those of  \cite{hor96}, lending credence to our evaluation of  how the gaugino condensate modifies the supersymmetry conditions.  This similarity can
 be extended in the following way. It was shown  in \cite{hor96} that it is natural to introduce a new four-form flux $\tilde{G}_{\it 4}$  of the form
\be
\tilde G_{\it 4}=G_{\it 4}+\langle S\rangle\, \delta^{\it 1}_{y^{11}=0}\wedge \omega_{\it 3}
\ee
where $\omega_{\it 3}$ is a three-form along the boundary $M$ and  $\delta^{\it 1}_{y^{11}=0}$ is the delta-function one-form (see footnote \ref{fn:delta}) localized at $y^{11}=0$ in the eleventh direction. It is natural to treat the new field $\tilde G_{\it 4}$ as a fundamental degree of freedom. In particular  $\tilde G_{\it 4}$ takes the  form of $G_{\it 4}$ calculated in the absence of the gauigino condensation.  This is an explicit example of the localized terms mentioned at the end of section \ref{sec:D5susy}. A very similar behavior will be discussed in section \ref{sec:gengaugino} in the context of the gaugino condensation on the D7-branes.


\subsection{Compactification and tadpole supersymmetry breaking}
\label{sec:susyb}

We have outlined the relation with the heterotic M-theory results in the case of non-compact internal space.
In the compact case, it was argued in \cite{hor96} that  the supersymmetry breaking occurs due to the non-local effects originating from the topological obstruction for finding global solutions of (\ref{Mhetkilling}).

In our setting we have the very same effect, which follows quite straightforwardly from (\ref{defomega_2}). Indeed, the r.h.s.\ defines a non-trivial second cohomology class while the l.h.s. is exact, which is not possible if the internal space is compact. Then, if $M$ is compact, supersymmetry is naturally broken and one expects the vacuum to destabilize, probably leading to a runaway behavior. Thus, we have a kind of `tadpole supersymmetry  breaking' of clear topological nature.

However, supersymmetry breaking is not unevitable  upon compactification, both in the heterotic M-theory \cite{hor96} (see also discussion in \cite{nilles}),
and in the type II case. Say, there are two stacks of the D5-branes on two isolated but homologous two-cycles $D_{1}$ and $D_2$ inside a CY three-fold.
We suppose for simplicity that there are no background fluxes in addition to those sourced by the D5-branes. Then, the equation (\ref{defomega_2})
is modified  to
\be\label{gcD5gen}
\d\Omega={2i\,\ell^6_s}\, \Big[\langle S_1\rangle \,\delta^{\it 4}_{D_1}+\langle S_2\rangle \,\delta^{\it 4}_{D_2}\Big]
\ee
where $\langle S_1\rangle$ and $\langle S_2\rangle$ are the gaugino condensates on the two stacks respectively. Since $D_1$ and $D_2$ are homologous, $\delta^{\it 4}_{D_1}$ and $\delta^{\it 4}_{D_2}$ define the same cohomology class and (\ref{gcD5gen}) can have a solution if $\langle S_1\rangle=-\langle S_2\rangle$. This conclusion has a clear four-dimensional interpretation. The generated low-energy superpotential for the CY K\"ahler moduli $t^I$ would be of the form $\calw_{\rm np}=A_1 e^{-b\alpha_1(t)}+A_2 e^{-b\alpha_2(t)}=c(\langle S_1\rangle +\langle S_2\rangle )$ and would thus identically vanish for $\langle S_1\rangle=-\langle S_2\rangle$.

The topological obstruction for the supersymmetric compactifications  may also be absent for more general configurations of D5-branes.  The condition is simply that the four-form $\sum_a \langle S_a\rangle \,\delta^{\it 4}_{D_a}$ governing the four-dimensional superpotential should be trivial as a cohomology class. Of course this is just one possible way to preserve supersymmetry in the case of compact internal space $M$  and flat four-dimensional external space $X_4$. We expect there should be other ways to preserve supersymmetry in the compact case without putting $\calw_{\rm np}=0$ of shell and allowing $X_4=$AdS$_4$.


\subsection{Relation with holographic  geometric transitions}
\label{sec:D5hol}

Returning to the non-compact case with $N$ D5-branes wrapping a rigid two-cycle $D$, consider the supersymmetry condition (\ref{defomega_2}) and compare it with the RR Bianchi identity
\be\label{F3BI}
\d F_{\it 3}= -\,\ell_s^2\, N\,\delta^{{\it 4}}_{D}\ .
\ee
The interpretation of (\ref{F3BI}) is that the D5-branes source the $F_{\rm 3}$ flux.
There is a similar interpretation of (\ref{defomega_2}). It is simply that the D5-branes, through the IR dynamics, source the holomorphic $(3,0)$-form or equivalently induce a deformation of
the complex structure. This point of view is illustrated by the following example
in the context of the gauge/gravity correspondence. Take the resolved conifold \cite{candelas} and wrap $N$ D5-branes on the blown-up two-sphere $S^2$ at the tip. In the large $N$ and near-horizon limit the backreacted background undergoes a geometric transition \cite{mn,vafa,vafa2,martellimalda}. The D5-branes disappear and the resolved conifold is replaced by a new background which satisfies the supersymmetry conditions from section \ref{sec:D5} with no localized sources.
The new background, of course, is the {\em deformed} conifold \cite{candelas}.  More explicitly, the original resolved two-sphere $S^2$ shrinks to zero size while a topologically  non-trivial three-sphere $S^3$ emerges. This transition includes the deformation of the complex structure, i.e. a deformation of the (3,0)-form $\Omega$ of the original CY. The role of the finite radius three-sphere at the tip of the conifold is two-fold. First, it supports the $N$ units of $F_{\it 3}$, which were originally sourced by the D5-branes according to (\ref{F3BI}). Second, from a purely holographic point of view, it regularizes the IR region of the geometry and is naturally associated to a mass gap, chiral symmetry breaking and gaugino condensation in the dual SYM theory.

\bigskip

There is a more precise way to relate $\Omega$ to the gaugino condensate of the dual gauge theory. One can compute the tension of a domain wall interpolating between the two nearby vacua of the dual $SU(N)$ SYM theory.  Namely, from the gauge-theory side we know that the domain wall between the two nearby vacua, in the large $N$ limit, is given by
\be\label{sugradwt}
T_{\rm DW}=2|\Delta W_{\rm np}|=2|W_{\rm np}(e^{\frac{2\pi i}{N}}-1)|\simeq \frac{4\pi}{N}\, |W_{\rm np}|=4\pi |\langle S\rangle |
\ee
where we have used the on-shell relation
\be\label{defs}
W_{\rm np}=N \langle S\rangle\ .
\ee
On the other hand, from the dual supergravity side, such a domain wall is represented by a D5-brane wrapping the minimal $S^3$ at the tip of the deformed conifold. Since $\Omega$ is the calibration for such a D-brane configuration \cite{lucal,thesis} from the supergravity side we get
\be\label{DWbranetension}
T_{\rm DW}=\frac{2\pi}{\ell_s^6}\Big|\int_{S^3}\Omega\Big|\ .
\ee
Hence, by comparing (\ref{sugradwt}) and (\ref{DWbranetension}) we get
\be\label{Speriod}
\Big|\int_{S^3}\Omega\Big|=2\ell^6_s\, |\langle S\rangle|\ ,
\ee
which fixes the relation between  $\langle S\rangle$ and $\int_{S^3}\Omega$ up to an unphysical overall phase.

\bigskip

Let us see now how this result is related to the interpretation of (\ref{defomega_2}) as the deformation of the complex structure sourced by the D5-branes.
Consider  $N$ D5-branes wrapping the blown-up two-sphere of the resolved conifold, i.e.\ the background {\em before} the geometric transition. We interpret (\ref{defomega_2}) as predicting that the  D5-branes supporting the gaugino condensate  `source' a deformation of the complex structure with a `strength' proportional to the gaugino condensate $\langle S\rangle$. One can measure the `flux' of sourced complex structure in the same way one would measure the RR-flux.  Let us recall that the geometry at the bottom of the resolved conifold can be identified with the $\calo(-1)\oplus\calo(-1)$ bundle over $\mathbb{P}^1\simeq S^2$.  Then we take a three-sphere  $S^3$ which surrounds $\mathbb{P}^1$ along a fiber. This $S^3$ is homologically trivial and indeed one can take a four-ball $B_4$ which stretches along the fiber and fills $S^3$, i.e.\ $\partial B_4=S^3$.  $B_4$ intersect $\mathbb{P}^1$ at one point. Then, by integrating (\ref{defomega_2}) on $B_4$ and using Stokes' theorem on the l.h.s.\ one arrives at
\be\label{Speriod2}
\int_{S^3}\Omega=2i\,\ell^6_s\, \langle S\rangle\ .
\ee
The conclusion is already clear: the non-trivial IR dynamics induces a geometric transition, in which a topologically trivial $S^3$ of the resolved conifold becomes a topologically non-trivial $S^3$ of the deformed conifold.  Then, our supergravity-based result (\ref{Speriod2}) precisely matches  the holographic result (\ref{Speriod}).

Let us also recall that the explicit solutions described in \cite{mn,martellimalda,buttietal} satisfy the tree-level supersymmetry conditions with no localized sources.
This is a further evidence that, at least  in the D5-brane case, there are no other corrections to the tree-level supersymmetry conditions besides those described by (\ref{defomega_2}).


\subsection{A comment on the case of fractional D3-branes}

In the above setting with the D5-branes wrapping the resolved conifold, one can  consider the limit when the
two-sphere $S^2$ at the tip of the cone is shrunk to the zero size. In this case the collapsed D5-branes  reduce to fractional D3-branes.
This is exactly the setting considered in \cite{ks} in the context of the gauge/gravity correspondence.
In this case, because of the conifold singularity, a treatment of the non-pertrubative dynamics on the D5-branes in terms of supergravity  may not be justified.
However, it is interesting to see what happens if we formally follow the steps outlined in the previous section in the case of the non-collapsed D5-branes.

First, the KS solution \cite{ks} is a subcase of the class of vacua considered in \cite{gkp}. In this case, the bulk superpotential
can be restricted to be the GVW one \cite{gvw}
\be\label{truncsupGVW}
\calw=-\frac{\pi}{\ell_s^8}\int_M \Omega\wedge \big(F_{\it 3}+ie^{-\phi}H\big) ,
\ee
while the gauge coupling associated to the gaugino condensate on $M$ fractional D3-branes is
\be\label{alphaKS}
\alpha= \frac{1}{\ell_s^2}\int_{D}(e^{-\phi}B-i C_{\it 2})\, .
\ee
By setting $\calt_{\it 2}=e^{-\phi}B-i C_{\it 2}$, we can write $\alpha\equiv\alpha(\calt)=\frac{1}{\ell_s^2}\int_{D}\calt_{\it 2}$
as in the case of the non-collapsing D5-branes.

Now all the steps followed for the case of the non-collapsing D5-branes can be repeated. Namely the equations (\ref{Fgaugino}) and (\ref{defomega_2}) still hold, although now the delta-function contribution $\delta^{\it 4}_D$ is not completely well defined as the cycle $D$ is shrunk to a zero size.   Although the zero size cycle may break the supergravity approximation, the prediction of (\ref{defomega_2}) still holds. Namely, as explained in section \ref{sec:D5hol} the geometry should develop a non-trivial three cycle, in full agreement with \cite{ks}.

Once the geometric transition has undergone, the two-cycle $D$ is the shrinking  $S^2$ of the deformed conifold.  Therefore   $\alpha$ is the difference of the two coupling constants  of the KS theory  \cite{KW,MP}.
As a result we obtain that the difference $\tau_{1}-\tau_{2}$ should be associated with the coupling of the $SU(M)$ theory on the D5-branes. One way to  check this is to
 compare the gaugino condensate of the $SU(M)$ theory, probed by the domain wall D5-brane wrapping $S^3$ at the tip of the deformed conifold (\ref{DWbranetension}), with  the expectation of ${\rm Tr}(\lambda_1\lambda_1-\lambda_2\lambda_2)$. Both quantities could be calculated using the low-energy superpotential $W(\tau_1,\tau_2)$ \cite{DKS}.


\section{The mirror conifold picture: condensing D6-branes}
\label{sec:D6branes}

Consider now  a mirror type IIA setting  in which a stack of D6-branes wraps a rigid three-cycle $\Sigma\subset M$ in the internal space. The D6-brane degrees of freedom reduce at low energies to a pure  four-dimensional SYM theory. The following steps will be practically identical to those followed in section \ref{sec:D5gen}, so we will proceed more quickly.

The string-frame metric has the form
\be\label{D6metric}
\d s^2_{X_4}=e^{2\phi/3}\d s^2_{X_4}+\d s^2_M\ .
\ee
Without taking into account the gaugino condensate,  the internal space has SU(3) structure, described by the pair $(J,\Omega)$. One of the supersymmetry conditions requires the internal space to be symplectic with the vanishing $H$-flux (for the complete set of conditions see \cite{GranaReview} and references therein)
\be\label{sympl}
\d J_{\rm c}=0\
\ee
where $J_{\rm c}:=J-iB$ is the complexified fundamental two-form.
This condition is mirror to (\ref{holsusy}) for the backgrounds with D5-branes. As in that case one can derive (\ref{sympl}) from a superpotential $\calw$ which depends on the SU(3) (or, in general, SU(3)$\times$SU(3)) structure of the internal space. Again, one should in principle consider the most general formula  for $\calw$ given below in (\ref{gensup}). However, for the current purpose it is sufficient to consider a truncation to the SU(3)-structure case
\be\label{D6sup}
\calw= \frac{\pi}{\ell^8_s}\int J_{\rm c}\wedge [F_{\it 4}+i\d (e^{-\phi}\Re\Omega)]\ .
\ee
 The relevant `chiral field' now is
\be
\calt_{\it 3}:= e^{-\phi}\Re\Omega-i C_{\it 3}
\ee
and the associated F-term is
\be\label{FtreeD6}
\calf_{\calt_{\it 3}}:= \frac{\delta \calw}{\delta\calt_{\rm 3}}\,= \, -\,\frac{i \pi }{\ell_s^8}\,\d J_{\rm c}\quad\qquad \text{(no gaugino condensate)}
\ee
so that the F-flatness condition $\calf_{\calt_{\it 3}}= 0$ is nothing but (\ref{sympl}).

Let us now consider the effect of the gaugino condensate along the lines of section \ref{sec:D5susy}. Since the D6-branes must wrap a special Lagrangian three-cycle $\Sigma$ \cite{lucal}, by performing the expansion as in (\ref{D5action}) it is easy to see that the holomorphic gauge coupling takes the form
\be\label{alphacaltII}
\alpha\equiv \alpha(\calt)= \frac{1}{\ell_s^3}\int_{\Sigma}(e^{-\phi}\Re \Omega-i C_{\it 3})\equiv \frac{1}{\ell_s^3}\int_{\Sigma}\calt_{\it 3}\ .
\ee
Analogously to (\ref{Fgaugino}), we now have
\be\label{FgauginoD6}
\calf_{\calt_{\it 3}}\,=\, \frac{\delta \calw}{\delta\calt_{\rm 3}}- 2\pi \, \langle S\rangle\,\frac{\delta \alpha}{\delta\calt_{\rm 3}}\,\equiv \, -\,\frac{i \pi }{\ell_s^8}\,\d J_{\it c}+\frac{2\pi}{ \ell_s^{3}}\, \langle S\rangle \,\delta^{\it 3}_{\Sigma}
\ee
which leads to the F-flatness condition
\be\label{defsym}
\d J_{\it c}= -{2i\,\ell^5_s}\, \langle S\rangle \,\delta^{\it 3}_{\Sigma}\ .
\ee
This is the IIA mirror counterpart of (\ref{defomega_2}). We then see that the effect of the gaugino condensation is to make the D6-branes act as  a localized ``source'' of the complexified symplectic structure. One could also  repeat, mutatis mutandis, the discussion of sections \ref{sec:hetM} and section \ref{sec:susyb}, reaching the same conclusions.

Finally, a concrete example is provided by the deformed conifold \cite{candelas} with $N$ D6-branes wrapping the three-sphere at the tip. This setting was originally discussed in \cite{vafa} (see also the uplift to  M-theory discussed in \cite{atyah}).  There, it was argued that the low-energy four-dimensional SYM theory on the D6-branes should be dual to a background symplectically equivalent to the resolved conifold, with the two-sphere supporting $N$ units of $F_{\it 2}$. Furthermore by computing the tension of a D4-brane domain wall wrapped on the resolved two-sphere $S^2$,  as in section \ref{sec:D5hol},  and comparing it with the expected gauge theory result one finds
\be\label{D6gaugino}
\Big|\int_{S^2}J_{\rm c}\Big|=2\,\ell^5_s\, |\langle S\rangle|\ .
\ee
This fixes the relation between the fundamental two-form and the gaugino condensate, up to an unphysical overall constant phase.
Repeating the arguments of section \ref{sec:D5hol}, it is easy to see that (\ref{defsym}) also leads to (\ref{D6gaugino}) with no reference to holography.


\section{Condensing D7-branes and generalized complex deformations}
\label{sec:D7}

Let us now turn to the case of  gaugino condensation localized on a stack of D7-branes. In absence of a gaugino condensate the backgrounds with the D7-branes are characterized by an integrable complex (actually, K\"ahler) structure and a holomorphic axion-dilaton $\tau=C_{\it 0}+ie^{-\phi}$.
There is a key difference between the D7 case and the cases with the D5 and D6 branes discussed previously. There, the gaugino condensation on a stack of the D5-branes (D6-branes) deformed the complex (symplectic) structure into a new complex (symplectic) structure.  As we will explain, a non-vanishing expectation value for the gaugino condensate on a stack of D7-branes has an effect of deforming the background complex structure into a {\em generalized} complex structure \cite{tentofour}.

For this reason it is practically unavoidable to use the explicit formalism of generalized geometry, from which we have refrained so far. Thus, we will review some basic facts about the use of generalized geometry in the context of flux compactifications, trying to introduce them in a self-consistent way.
A comprehensive review of this subject, including  a  list of references, can be found in \cite{paulreview}.

\subsection{Intermezzo I: $\caln=1$ vacua and generalized geometry}
\hlt{}

\label{sec:genback}

We start with  the IIA/IIB backgrounds of the form $X_{10}=X_4\times M$, with a string-frame metric
\be\label{genmetric}
\d s^2_{X_{10}}=e^{2A}\d x^\mu\d x_\mu+\d \hat s^2_M\, ,
\ee
general dilaton $\phi$, $H$-flux (which can be locally written as $H=\d B$) and internal RR-fluxes
\be
F=\sum_{\it k} F_{\it k}\qquad\qquad {\it k}=\left\{ \begin{array}{lll} {\it 0,2,4,6} \quad &{\rm IIA} \\ {\it 1,3,5} \quad & {\rm IIB}\end{array}\right.
\ee
Locally, away from localized sources, we can write $F= \d _H C$, where $C=\sum_{\it k}C_{\it k}$ (${\it k}$ odd/even in IIA/IIB) is the RR potential.\footnote{More precisely, in IIA with $F_{\it 0}\neq 0$, one has $F=e^{-B}F_{\it 0}+\d_H C$.} For future convenience, we have decided to denote the string-frame internal metric with $\d \hat s^2_M$. This notational choice is due to the possible existence of a more natural  metric $\d s^2_M$, conformally related to $\d\hat s^2_M $. In the following, we will distinguish quantities involving the complete string-frame metric with a hat $\hat{}$. For instance, by $\hat *$ we will mean the six-dimensional Hodge-start operator computed with $\d \hat s^2_M$. Hence  the symbol $*$ will be reserved for the Hodge-operator computed using  the conformally-rescaled internal metric $\d s^2_M$.

Now, a key point is that the complete information about $\d \hat s^2_M$, $\phi$, $e^A$, $B$-field and two arbitrary normalized chiral spinors
in six dimensions $\eta_{1,2}$ (with equal norm $\eta_1^\dagger\eta_1=\eta_2^\dagger\eta_2=: |a|^2$)
can be encoded in two complex polyforms $\calz$ and $T$, defined by the Clifford map as follows\footnote{The Clifford map is one-to-one map between a polyform $\omega=\sum_{\it k}\omega_k$ on $M$ and the matrix with spinorial indices  $\slashed{\omega}:= \sum_{\it k}\frac{1}{\it k!}\omega_{m_1\ldots m_{\it k}}\gamma^{m_1}\cdots\gamma^{m_{\it k}}$.}
\be\label{defps}
e^{-B}\wedge \calz\ \leftrightarrow\ -\frac{8i}{|a|^2}e^{3A-\phi}\eta_1\otimes\eta_2^T\quad ,\quad\quad e^{-B}\wedge T\ \leftrightarrow\ -\frac{8i}{|a|^2}e^{-\phi}\eta_1\otimes\eta_2^\dagger\ ,
\ee
where
\be
e^{B}\wedge:= (1+B+\frac12 B\wedge B+\frac{1}{3!}B\wedge B\wedge B)\wedge \ .
\ee
For a reason which will becomes obvious in a moment we choose $\eta_{1},\eta_{2}$ to be of opposite chirality in case of IIA and the same chirality in case of IIB theory.
Then $\calz$ is even/odd and  $T$ is odd/even respectively:
\be
\label{rankexp}
\text{IIA:}\ \left\{\begin{array}{l}  \calz=\calz_{\it 0}+\calz_{\it 2}+\calz_{\it 4}+\calz_{\it 6} \\
T=T_{\it 1}+T_{\it 3}+T_{\it 5}\end{array}\right.\, ,\quad \text{IIB:}\ \left\{\begin{array}{l}  \calz=\calz_{\it 1}+\calz_{\it 3}+\calz_{\it 5} \\
T=T_{\it 0}+T_{\it 2}+T_{\it 4}+T_{\it 6}\end{array}\right.\, ,
\ee
Defined as in (\ref{defps}), $\calz$ and $T$ turn out to be  pure spinors of the $T_M\oplus T_M^*$ bundle  and obey a compatibility condition\footnote{Technically speaking, the conditions (\ref{compcond}) are equivalent to demanding that $\calz$ and $T$ define an SU(3)$\times$ SU(3)-structure on $T_M\oplus T^*_M$, which can be in turn associated to the existence of the two internal spinors $\eta_{1,2}$, each defining an SU(3) structure on $T_M$.}  (see e.g.\  \cite{paulreview} for more details)
\bea\label{compcond}
&&\langle\iota_v\calz, T\rangle=\langle\iota_v\bar\calz, T\rangle=0\, ,\qquad\quad\quad\forall\ v\in T_M\ ,\cr
&& \langle\chi\wedge \calz,T\rangle=\langle\chi\wedge \bar\calz,T\rangle=0\, ,\quad\quad\forall\ \chi \in T^*_M\ .
\eea
Here we used the Mukai  pairing defined for the two arbitrary polyforms $\calz,T$
\bea
\langle\calz, T\rangle=[\calz\wedge \sigma(T)]_{\it 6}
\eea
and $\sigma$ is the involutive operator that reverses sign in front of some forms $\sigma (\omega_k)=(-)^{\frac{k(k-1)}{2}}\omega_k$. Basically, $\sigma$ reverses the order of the indices
\be
\sigma(\d y^{m_1}\wedge \ldots\wedge \d y^{m_k})=\d y^{m_k}\wedge \ldots\wedge \d y^{m_1}\ .
\ee
The conditions (\ref{compcond}) are not only necessary but sufficient. If we take a pair of pure spinors $\calz$,$T$  on $T_M\oplus T_M^*$ such that  (\ref{compcond}) is satisfied, then there are $\eta_{1,2}$, metric, $B$ field, dilaton and warping such that $\calz$ and $T$ can be written in the form (\ref{defps}).

Now, if the background respects supersymmetry the spinors $\eta_{1,2}$ can be identified with internal components of  the ten-dimensional Killing spinors
\be
\epsilon_1=\zeta\otimes \eta_1+\text{c.c.}\, ,\quad \epsilon_2=\zeta\otimes \eta_2+\text{c.c.}
\ee
where $\epsilon_{1,2}$ are the MW spinors in ten-dimensions and $\zeta$ is an arbitrary constant Weyl spinor in four dimensions. Since $\calz$ and $T$ contain full information about  $g$, $\phi$, $B$, $ A$ and $\eta_{1,2}$   the SUSY conditions can be written in terms of these two objects, with  only RR-fields  $F$ as external ingredients. Indeed, it has been shown in \cite{gmpt} that, for Minkowski $X_4$, the background supersymmetry conditions can be rewritten in the form
\bseq\label{susycond}
\begin{align}
\d\calz&=0 \label{susycond1}\ ,\\
\d(e^{2 A}\Im T)&=0\label{susycond2}\ ,\\
 \d(e^{4 A}\Re T)&=e^{4 A}e^{B}\wedge \hat*\, \sigma( F)\ . \label{susycond3}
\end{align}
\eseq
It is important to stress that (\ref{susycond1}) implies that $M$ has an integrable  generalized complex structure defined by $\calz$. Roughly speaking, this is equivalent to the definition, on local patches, of hybrid complex-symplectic coordinates.

For the following, it is important to observe that (\ref{susycond3}) can be written in the alternative form
\be
 \d(e^{4 A}\Re T)=e^{B}\wedge \tilde F\ . \label{susycond3dual}
\ee
where $\tilde F$ is the polyform containing the RR-fluxes dual to $F$, according to the relation
$\d^4 x\wedge \tilde F=*_{10}\sigma(F)$, i.e.
\be\label{defdualF}
\tilde F=e^{4A}\,\hat*\, \sigma( F)\, .
\ee
Because $\tilde F$ is $\d_H$-closed we can locally write $\tilde F=\d_H \tilde C$ and it is the combination $\d^4x \wedge \tilde C$ that couples electrically to the space-filling D-branes and O-planes. A crucial observation is that (\ref{susycond3dual}) admits an interpretation as a condition for $e^{4A}\Re T$ to be a calibration for the space-filling D-branes \cite{lucal}.
The calibration structure guarantees classical stability of branes. Hence we regard  the form (\ref{susycond3dual}) to be more  fundamental  than (\ref{susycond3}).
Of course, for purely bosonic backgrounds the two conditions are completely equivalent. However, as we will see, this will no longer be the case  in the presence of the non-vanishing gaugino condensates.
Indeed, a similar phenomenon occurs in the case of the heterotic string backgrounds \cite{Held:2010az}.

In addition to SUSY constraints (\ref{susycond1}-\ref{susycond3}) one needs to impose the BI for the $H$-flux, $\d H=0$, and for the internal RR-fluxes $F$ (or, equivalently, the equation of motion for $\tilde F$). The latter can be written in a compact polyform notations
\be\label{RRbi}
\d_H F\equiv \d F+H\wedge F=-e^{-B}\wedge j
\ee
where $j$ contains the RR localized sources
\be\label{RRcurrent}
j\equiv j_{\rm D}+j_{\rm O}:=\!\!\!\!\!\!\sum_{a\in \text{D$p$-branes}}\ell_s^{7-p}\,\sigma(\delta^{\it 9-p}_{D_a})\wedge e^{-2\pi\alpha^\prime {\rm F_a}}\ -\!\!\!\!\!\!\sum_{b\in \text{O$q$-planes}}2^{q-5}\ell_s^{7-q}\,\sigma(\delta^{\it 9-q}_{O_b})\ .
\ee
In (\ref{RRcurrent}), $D_a$ are the cycles wrapped by the D-branes, $O_b$ are the ones wrapped by the orientifolds and ${\rm F_a}$ is the gauge flux supported on $D_a$.\footnote{For simplicity, we do not include curvature corrections. Moreover our notations assume  only U(1)-bundles supported on $D_a$. This can be generalized to the non-abelian bundle configurations.}  In the presence of the orientifolds, $M$ has to be considered as the covering space.  Furthermore, in order to preserve supersymmetry the localized sources have to satisfy the proper calibration conditions.

The  equations of motion for the internal RR-fluxes $F$,
\be
\d_H\tilde F=\d_H[e^{4{A}}\,\hat *\sigma(F)]=0\label{RREoM}
\ee
 follow from the supersymmetry condition (\ref{susycond3}) due to $\d_H^2=0$.  The equation of motion for the $H$-field
\be\label{HEoM}
\d(e^{4A-2\phi}\,\hat *H)-e^{4A}\sum_k \hat * F_{\it{k+2}}\wedge F_{\it k}=-\Big[e^{4A}\,\Re T\wedge\sigma(j_{\rm D})\wedge e^{B}\big]_{\it 4}
\ee
follows from the supersymmetry conditions as well \cite{pauldimi}.

As an illustration, let us revisit the case with the D5-branes discussed in section \ref{sec:D5gen}. In this case the metric is of the form (\ref{D5metric}),  hence we can identify $\d \hat s^2_M$ with $\d s^2_M$ and $e^{2A}=e^\phi$. The  pure spinors take the form
\be\label{D5ps}
\begin{array}{l}  \calz=\Omega \\
T=-ie^{-\phi}e^{iJ+B} \end{array}
\qquad\qquad \text{backgrounds with D5-branes}
\ee
On the other hand, for the classical backgrounds with the D6-branes of section \ref{sec:D6branes}, we identify $\d \hat s^2_M$ from (\ref{genmetric}) with $\d s^2_M$ in (\ref{D6metric}) and set $e^{A}=e^{\phi/3}$. The on-shell pure spinors are
\be\label{D6ps}
\begin{array}{l}
\calz=e^{iJ+B} \\
T=e^{-\phi}\Omega \end{array}
\qquad\qquad \text{backgrounds with D6-branes}
\ee
In both cases the pair $(\Omega,J)$ defines the SU(3)-structure and satisfies the normalization condition
\be\label{jomega}
-\frac{i}{8}\,e^{k\phi}\, \Omega\wedge \bar\Omega=\frac{1}{3!}\, \,J\wedge J\wedge J=\text{vol}_M\ .
\ee
where $\text{vol}_M=\d^6 y\sqrt{\det g_M}$ and $k=-1$ for the D5-brane and $k=0$ for the D6-brane background.

With the choice (\ref{D5ps}), the condition (\ref{susycond1}) implies the integrability of the almost complex structure defined by $\calz\equiv \Omega$. On the other hand, in the case of (\ref{D6ps}) the condition  (\ref{susycond1}) implies the existence of a symplectic structure defined by $J$. This is exactly what we saw in sections \ref{sec:D5gen} and \ref{sec:D6branes}.


\subsection{Unperturbed backgrounds with D7-branes}
\label{sec:D7back}

After the machinery of generalized geometry is introduced
we are back to the backgrounds with the D7-branes (and possibly the O7-planes).
In what follows we assume that there is no localized D3-brane charge,\footnote{This also means that we set to zero the world-volume fluxes on the D7-branes. Moreover, we neglect possible D3-brane charge induced on the D7-branes by the curvature terms.} O3-branes or non-trivial five-form as well as the 3-form fluxes, unless specified.

The ten-dimensional space is a product $X_{10}=X_4\times M$ of the  four-dimensional Minkowski space $X_{4}$ and  the internal six-dimensional space $M$. The D7-branes (and possibly the O7-planes)
filling $X_4$ are wrapping some internal four-cycles $D_a$.
In the {\em string frame}, the metric takes the following form
\be\label{Ewarpedmetric}
\d s_{X_{10}}^2 =e^{\phi/2}\,(\d x^\mu\d x_\mu+\d s_M^2)
\ee
where the dilaton $\phi$ depends only on the coordinates along $M$. Hence in this case the internal metric is $\d \hat s^2_M=e^{\phi/2}\d s^2_M$ and the warping is $e^A=e^{\phi/4}$. There is no RR or $H$-flux and the pure-spinors have the  form
\be\label{psD7}
\begin{array}{l}  \calz_{\text{F-th}}=\Omega \\
T_{\text{F-th}}=e^{-\phi}\exp(i\, e^{\phi/2}J) \end{array}
\qquad\qquad \text{(backgrounds with D7-branes)}
\ee
where the suffix ``F-th" indicates explicitly that we are working with the F-theory-like backgrounds, i.e.  backgrounds with a holomorphic axion-dilaton which admit a supergravity description. In (\ref{psD7}) the pair $(\Omega,J)$ defines an SU(3)-structure associated to the metric $\d s^2_M$, and are normalized as follows
\be\label{normjomega}
-\frac{ie^{-\phi}}{8}\Omega\wedge \bar\Omega=\frac{1}{3!}\, \,J\wedge J\wedge J=\,\text{vol}_M\ .
\ee
Here $\text{vol}_M=\d^6 y\sqrt{\det g_M}$ is the volume form computed with the metric $\d s^2_M$.
From these restrictions and general conditions reviewed in section \ref{sec:genback}, it is possible to derive the remaining
properties of these backgrounds. The supersymmetry condition (\ref{susycond1}) implies that $\calz_{\text{F-th}}$ defines an integrable complex structure. The metric $\d s_M^2$ in (\ref{Ewarpedmetric}) is K\"ahler, with K\"ahler form $J$ and the holomorphic $(3,0)$-form $\Omega$. The dilaton $\phi$ combines with the RR zero form $C_{\mathit{0}}$ into a holomorphic axion-dilaton $\tau:=C_{\mathit{0}}+ie^{-\phi}$
\be
\delbar\tau=0\ ,\quad \delbar:=\d \bar z^{\bar\imath}\wedge \delbar_{\bar\imath}\ .
\ee
 The D7-branes and the O7-planes act as the localized source
\be
j=-\!\!\!\!\!\!\!\sum_{a\in \text{D7-branes}}\delta^{\it 2}_{D_a} + 4\!\!\!\!\!\!\!\sum_{b\in \text{O7-planes}}\delta^{\it 2}_{O_b}\ .\label{jd7}
\ee
in the Bianchi identity for $F_\itone=\d C_\itzero$
\be\label{tauBI}
\d F_\itone =  2i\del\delbar e^{-\phi}=\delbar\del\tau=-j\ .
\ee

Furthermore, the metric $\d s^2_M$ satisfies the Einstein equation
\be\label{dileinst}
R_{i\bar\jmath}=\nabla_i\bar\nabla_{\bar\jmath}\,\phi\ .
\ee
Here, and in what follows  $x^\mu$ ($\mu=0,\ldots,3$) are the coordinates along $X_4$, $y^m$ ($m=1,\ldots,6$) are the real coordinates along the internal space $M$, and $z^i$ (i=1,2,3) are the complex coordinates on $M$.
Eventually the holomorphic $(3,0)$ form $\Omega$ satisfies the following equations
\be\label{diffomega}
\bar\nabla_{\bar\imath}\Omega=0\ ,\qquad\qquad \nabla_{i}\Omega=(\nabla_i\phi)\, \Omega\ .
\ee
Clearly, when $\tau$ is constant the internal K\"ahler metric $\d s^2_M$ is Ricci flat and one recovers the standard case of the Calabi-Yau manifold $M$.


\subsection{Gaugino condensation and generalized geometry}
\label{sec:gengaugino}

We would like to study the effect of gaugino condensation on a stack of D7-branes. To  engineer a D-brane configuration
which exhibits such a behavior in the IR is not as straightforward as in the cases with the D5 and D6 branes. The consistent configurations of D7-branes wrapping compact divisors are subject to constraints coming from tadpole cancelation conditions (see e.g.\ \cite{cacha01}  and the  recent discussion in \cite{franco10}). Hence one has to consider a combination of (intersecting) the D7-branes and the O7-planes. The intersections generically alleviate the topological obstructions but also give rise to light fields prone to complicate the low-energy dynamics.
However, in a general case one can turn on fluxes through the intersection, uplifting the massless  modes. The resulting low-energy theory on a given divisor is SYM with some massive matter.
In the following,  we will simply assume that gaugino condensation  does indeed occur, without specifying any details about its dynamical origin.

We will see shortly that the  gaugino condensate on D7-branes will change the supersymmetry condition in such a way that it can not be solved by $\calz_{\text{F-th}}$ associated with the ordinary complex structure given by (\ref{psD7}). Rather it will be  deformed into a new $\calz$ associated with a {\em genuine} generalized complex structure.
Therefore we go back to the most general setting of section \ref{sec:genback}, and re-derive how the  F-flatness condition (\ref{susycond3}) is modified by a gaugino condensate on a stack of the D-branes \cite{tentofour}, but along the lines followed in sections \ref{sec:D5gen} and \ref{sec:D6branes}.

\subsection{Intermezzo II: $\langle S\rangle$-deformed condition for $\calz$, in general}

As argued in \cite{tentofour}, the general IIA/IIB classical supersymmetry conditions (\ref{susycond}) can be obtained from the superpotential
\be\label{gensup}
\calw=\frac{\pi}{\ell_s^8}\int_M\langle \calz, F^{\rm tw}+i\d\Re T  \rangle
\ee
where we introduced the twisted RR-fields
\be
F^{\rm tw}:=e^B\wedge F\qquad\Rightarrow\quad \d F^{\rm tw}= -j\ .
\ee
The reader can check that by plugging  (\ref{D5ps}) and (\ref{D6ps}) into (\ref{gensup}) one gets (\ref{truncsupD5}) and (\ref{D6sup}), up to unimportant overall constant phases, which can be reabsorbed into a definition of $\calz$.

The condition (\ref{susycond1}) is obtained by extremizing $\calw$ with respect to the `chiral field'
 \be
 \calt:=\Re T-i C^{\rm tw}\ ,
 \ee
 where $C^{\rm tw}$ is locally defined by $F^{\rm tw}=\d C^{\rm tw}$. Indeed, the associated F-term is\footnote{See \cite{tentofour} for the precise definition of complex structure on the $\calt$ field space, which is implicit in the definition of the F-term.}
\be
\calf_{\calt}=\frac{\delta \calw}{\delta \calt}=\frac{i\pi}{\ell_s^8}\d \calz\quad\qquad \text{(no gaugino condensate)}
\ee
where we have introduced the  functional derivative defined through the  Mukai pairing: $\delta_\calt\calw=\int_M\langle \frac{\delta\calw}{\delta\calt},\delta\calt\rangle$. The F-flatness condition is simply $\calf_\calt= 0$ which immediately  reproduces (\ref{susycond1}). The remaining conditions in (\ref{susycond}), including their extension to the AdS case, can be also derived from $\calw$ using an appropriate K\"ahler potential.

Let us now add a gaugino condensate on a stack of  supersymmetric D$p$-branes, working in approximation of local model, i.e.\  in the case of non-compact $M$, as in sections \ref{sec:D5gen} and \ref{sec:D6branes}. One can repeat the steps of section \ref{sec:D5susy} almost verbatim. By using the fact that the D$p$-branes preserve supersymmetry and must be calibrated, the associated  four-dimensional complexified coupling constant obtained by dimensional reduction of the D$p$-brane action  wrapping a cycle $D$, possibly supporting a world-volume  flux F, is
\be\label{holgc}
\alpha(\calt)=\frac{1}{\ell_s^{p-3}}\int_D \calt\wedge e^{2\pi\alpha^\prime {\rm F}}\equiv \frac{1}{\ell_s^4}\int_M\langle\, \calt, j_{\rm np}\rangle
\ee
with $j_{\rm np}=\ell_s^{7-p}\,\sigma(\delta^{(9-p)}_{D})\wedge \exp(-2\pi\alpha^\prime {\rm F})$. The difference between $j_{\rm np}$ and $j$ defined in (\ref{RRcurrent}) is that $j$ includes all D7 (and O7) branes while $j_{\rm np}$ only includes those with the non-trivial gaugino condensate.  By adapting  (\ref{modFterms}) to the present setting one gets the following general expression for the F-term in the presence of the gaugino condensate
\be\label{Fgauginogen}
\calf_{\calt}=\frac{\delta \calw}{\delta \calt}- 2\pi \, \langle S\rangle\,\frac{\delta \alpha}{\delta\calt}\,\equiv\frac{i\pi}{\ell_s^8}\d \calz+\frac{2\pi}{\ell_s^4}\langle S\rangle\, j_{\rm np}\ .
\ee
This is a generalization of (\ref{Fgaugino}) and (\ref{FgauginoD6}). Then, from the F-flatness condition $\calf_\calt=0$ one obtains the following modification of the supersymmetry condition (\ref{susycond1})
\be\label{modsusycond}
\d\calz=2i\,\ell_s^4\, \langle S\rangle\, j_{\rm np}\ .
\ee

Equation (\ref{modsusycond}) was first obtained in \cite{tentofour}, following a formally equivalent but conceptually different derivation, by modifying the tree-level superpotential (\ref{gensup}) by adding a non-perturbative superpotential depending on $\calt$. The form of this non-perturbatively generated superpotential was guessed based on the form of superpotential in the four-dimensional theory. On the other hand, the logic followed here is completely ten-dimensional. Our results are manifestly independent of the details of the effective four-dimensional theory.

\subsection{Back to D7-branes with $\langle S\rangle\neq 0$}

We are now ready to understand the main difference between the case of condensing D7-branes and the previously discussed cases with D5 and D6-branes. Suppose there is a stack of the condensing D7-branes, wrapping a four-cycle $D$ with the vanishing  world-volume flux F. In this  case (\ref{modsusycond}) reads
\be\label{D7def}
\d\calz=-2i\,\ell_s^4\, \langle S\rangle\, \delta^{\it 2}_D \qquad\qquad\text{(\,D7-branes with $\langle S\rangle\neq 0$)}
\ee
It is now clear  that  (\ref{psD7}) is not compatible with (\ref{D7def}), and $\calz$ must necessarily have  a one-form contribution $\calz=\calz_{\it 1}+\calz_{\it 3}$, with $\calz_{\it 1}\neq 0$. This implies that $\calz$ defines a genuinely {\em generalized} complex structure. Thus, the gaugino condensate on the D7-branes catalyzes a  deformation outside the realm of ordinary complex geometry. This was one of the main results following from the equation (\ref{D7def}).

The fate of the remaining supersymmetry conditions (\ref{susycond2}) and (\ref{susycond3}) was not clarified in \cite{tentofour}. The derivation above suggests that they should be unmodified. However, this conclusion could be too naive and may not take into account a possible subtlety related to the presence of the RR-fluxes in the superpotential (\ref{gensup}). Indeed, experience with the heterotic M-theory \cite{hor96}  tell us that, in the presence of the gaugino condensate the flux appearing in the equations of motion and the supersymmetry conditions naturally combine with a singular form proportional to the gaugino condensate and localized on the defect where the nonperturbative effect is taking place (see also \cite{drsw,Frey} for the weakly-coupled heterotic counterpart of this effect). This was already mentioned  in section \ref{sec:hetM}. Similarly in our case it is natural to allow for a possible correction to the fluxes of the schematic form
\be\label{fluxcorrection}
\text{(flux)}\quad \longrightarrow\quad \text{(flux)}+\langle S\rangle\times \text{(localized term)}\ .
\ee
 This issues will be considered in detail in the following section, where we also clarify the relation with related observations of \cite{bau10}.


\section{Condensing D7-brane back-reaction: First order deformation}
\label{sec:firstord}

We now address the problem of identifying the supersymmetric solution that represents the backreaction of the gaugino condensate  $\langle S\rangle\neq 0$ on the D7-branes wrapping the cycle $D$. Besides the D7-branes on $D$ we assume that there are other D7-branes, indicating with  $\{D_a\}$ the complete set of wrapped divisors. For simplicity we assume there are no O7-planes in the set-up. The case with the O7-planes will be discussed in the end of the section (\ref{sec:duality}).

The equation (\ref{D7def}) admits an exact solution
\be\label{newcalz}
\calz=\theta+\Omega
\ee
where $\theta$ is a $(1,0)$-form subject to the conditions
\be\label{thetadef}
\partial\theta=0\, ,\quad\quad \delbar \theta=-2 i\,\ell_s^4\,\langle S\rangle\,\delta^{\it 2}_D\ .
\ee
As a consequence, the divergence of $\theta$ is given by a delta-like term, localized on $D$
\be\label{nablatheta}
\nabla^m\theta_m= -2\,\ell_s^4\,\langle S\rangle\, \delta^{(0)}_{D}\ .
\ee
Here
\be
\delta^{(0)}_D=-ig^{i\bar j}\delta^{\it 2}_{D\,i\bar j}\ ,
\ee
is a scalar delta-function localized on $D$.

The first condition in (\ref{thetadef}) can be locally integrated by
\be\label{loctheta}
\theta=-\frac{\ell_s^4}{\pi}\,\del w\, ,
\ee
where a holomorphic $w$ can be identified with the superpotential experienced by a probe D3-brane \cite{lucasup}. As we will see later, such an   identification  is completely consistent with the form of the D3-brane superpotential from the IASD fluxes discussed in \cite{bau10}. Now, the second condition in (\ref{thetadef}) written in a local patch becomes  $\delbar \del w=2\pi i\,\langle S\rangle\,\delta^{\it 2}_D$. This equation can be  immediately integrated using the Poincar\'e-Lelong lemma (see e.g.\ \cite{GH})
\be\label{intw}
w(z)=\langle S\rangle\,\log h(z)+w_0
\ee
where  $h(z)$ is the holomorphic section of the divisor line bundle $\call_D$ that defines $D$ through the equation $h(z)|_D=0$, and $w_0$ plays the role of the  integration constant.

Having solved  (\ref{D7def}), we now turn to consider the remaining supersymmetry conditions. In the absence of a gaugino condensate, they are given by (\ref{susycond2}) and (\ref{susycond3}). In section (\ref{sec:genback}) we introduced the alternative form (\ref{susycond3dual}) of (\ref{susycond3}) that has a simple physical interpretation in terms of the calibration condition.
We regard this form as the more fundamental and impose the following two supersymmetry conditions
\bseq
\begin{align}
\d(e^{2 A}\Im T)&=0\qquad\qquad\qquad\ \,(\text{Condition II})\label{defsusycond2} \\
 \d(e^{4 A}\Re T)&=e^{B}\wedge \tilde F\qquad\qquad(\text{Condition III})\label{defsusycond3}
\end{align}
\eseq
We will see that when $\langle S\rangle\neq 0$, (\ref{defsusycond3}) is in fact not equivalent to (\ref{susycond3}).

In the following we will look for a perturbative  solution of the supersymmetry equations deformed by the effect of the gaugino condensate, with $\langle S\rangle$ playing the role of the expansion parameter. For instance, this should be natural for small bare `t Hooft coupling on the condensing D7-brane. We then expand the pure spinors  in powers of  $\langle S\rangle $ (not to be confused with the rank expansion (\ref{rankexp}))

\bseq\label{psexp}
\begin{align}
\calz&= \calz^0+\calz^1+\ldots \label{psexp1}\ \\
T&= T^0+T^1+\ldots\label{psexp2}\
\end{align}
\eseq
where we identify $\calz^0$ and $T^0$ with $\calz_{\text{F-th}}$ and $T_{\text{F-th}}$ of (\ref{psD7}), while $\calz^1$ and $\calt^1$ are first-order in $\langle S\rangle$. We have already provided above an expression for $\calz$ which is first order in $\langle S\rangle$ and solves the condition (\ref{D7def}) exactly. We then need to consider $T$ and the remaining conditions (\ref{defsusycond2},\ref{defsusycond3}).


\subsection{First step: supersymmetry condition II}
\label{sec:Dterm}

So far we have a deformation of $\calz$   (\ref{newcalz}) that solves (\ref{D7def}).We now ask if and how it is possible to deform the other pure spinor $T$ such that it is compatible with $\calz$ (\ref{compcond})  and satisfies the remaining conditions. In particular,  in this section we focus our attention on (\ref{defsusycond2}).
 As we now show, there is a simple way to identify such a deformation.\footnote{Of course, there could be other deformations of $T^0$ into $T$, but they should correspond to the ordinary classical deformations of the original background not related to the non-perturbative effect.}

In order to proceed, it is convenient to recast (\ref{newcalz}) as the result of  the so called {\em holomorphic $\beta$-deformation} of $\calz^{0}\equiv\Omega$. It is described by a real bivector
\be
\beta=\beta^{2,0}+\beta^{0,2}=\frac12\beta^{ij}(z)\frac{\del}{\del z^i}\,\wedge \frac{\del}{\del z^j}+\text{c.c.}
\ee
whose $(2,0)$-component $\beta^{2,0}(z)$ is holomorphic. The action of $\beta$ on a spinor  can be written as $e^{\iota_\beta}$. In particular
\be\label{betacalz}
\calz=e^{\iota_\beta}\calz_0=\iota_\beta\Omega+\Omega\ ,
\ee
where $\iota_\beta\Omega:=\frac12\beta^{mn}\Omega_{mnp}\d y^p=\frac12\beta^{ij}\Omega_{ijk}\d z^k$. By comparing (\ref{betacalz}) and (\ref{newcalz}), we see that  $\beta$ is completely determined by
\be\label{deftheta}
\iota_\beta\Omega\,=\,\theta\ .
\ee
Notice that such $\beta^{2,0}$ with $\theta$ satisfying (\ref{thetadef}) automatically defines a holomorphic Poisson structure, $\beta^{i[j}\partial_i\beta^{lk]}=0$.

Now, the very same $\beta$-deformation can be applied to the pure spinor $T^0\equiv T_{\text{F-th}}$ to produce a new pure spinor $T$, which will be automatically compatible with $\calz$. For the time being, we work to first order in $\langle S\rangle$ i.e. in $\beta$. A discussion of higher order effects of the $\beta$-deformation can be found in appendix \ref{sec:allorder}. At first order the $\beta$-deformed $T$ is given by
\be\label{1orderT}
T=T^0+\iota_\beta T^0 \ .
\ee
Recall that the pair $\calz$ and $T$ specify the complete information about the internal metric $g$, the $B$ field,  dilaton $\phi$ and the warp factor $e^A$.  By direct inspection of (\ref{betacalz}) and (\ref{1orderT}), it turns out that the metric, the dilaton and the warp factor  are unmodified. Only the $B$-field is deformed as follows
\be\label{Bfield}
 B= \hat{g}^0\,\beta\, \hat{g}^0 ={{e^\phi}\over 2}\iota_\beta(J\wedge J)= \frac14\Re(\bar\theta\lrcorner \Omega)
\ee
with $\bar\theta\lrcorner \Omega\equiv \bar\theta^m\iota_m\Omega$. The direct derivation of this result is somewhat technical; some details of this calculation can be found in the appendix   \ref{sec:allorder}. From (\ref{Bfield}) one can easily find the following explicit expression for the $H$-field
\be\label{defHflux}
H=\frac14\Re(\bar\del\bar\theta^m\wedge \iota_m\Omega)+\frac14\Re\big[(\partial_m\phi)\bar\theta^m\Omega\big]+ \frac14 \Re[\nabla^m\bar\theta_m\Omega]\ .
\ee

Coming back to (\ref{defsusycond2}), we need the following first order identity
\bea\label{imt}
e^{2 A}\Im T&=&e^{\phi/2}\Im T^0-\frac{1}{3!}\,e^{\phi}\iota_\beta(J\wedge J\wedge J)+\ldots\cr
&=& (J-{e^{\phi}\over 3!}J\wedge J\wedge J)+\frac{i}{8}\, \iota_\beta(\Omega\wedge \bar\Omega)+\ldots\cr
&=&(J+{i\over 8}\Omega\wedge \bar \Omega)+\frac{1}{4}\, \Im(\bar\theta \wedge \Omega)+\ldots\ ,
\eea
It is now easy to see that  (\ref{defsusycond2}) is indeed satisfied because of  $\d\Omega=0$ and  $\del\theta=0$.

It now remains to discuss (\ref{defsusycond3}). In order to clarify its meaning, we have to understand how the gaugino condensate modifies the duality relation (\ref{defdualF}). We focus on this problem in the next section.


\subsection{Second step: supersymmetry condition III and flux duality}
\label{sec:duality}

Now we use the condition (\ref{defsusycond3}) to define $\tilde{F}$. Obviously this will guarantee that (\ref{defsusycond3}) is satisfied.   From
\be\label{reT}
\Re T =e^{-\phi}-\frac12\,\iota_\beta(J\wedge J)-\frac12\,  J\wedge J+\ldots=e^{-\phi}-\frac14\, e^{-\phi}\,\Re(\bar \theta\lrcorner\Omega)-\frac12\, J\wedge J+\ldots
\ee
and using (\ref{Bfield}), it is easy to see that, to first order in $\beta$, (\ref{defsusycond3}) implies
\bseq
\begin{align}
\tilde{F}_{\it 1}&=0\ , \\
\tilde F_{\it 3}&=-H\ ,\\
\tilde F_{\it 5}&=-\frac {e^{\phi}} 2\d\phi\wedge J\wedge J\ .
\end{align}
\eseq
Our next step would be to calculate $F$. We can not  use (\ref{defdualF}) because, as was  mentioned earlier we expect the relation between $F$ and $\tilde{F}$ is modified by the local terms proportional to $\langle S\rangle\neq 0$. To find the correct relation one has to look at the full action in the bulk modified by the local terms on the  D7-branes. It turns out that the original relations $\tilde F_{\it 1}=*F_{\it 5}$ and $\tilde F_{\it 5}=e^{2\phi}*F_{\it 1}$ are unmodified. Hence $F_{\it 5}=0$ and the axion-dilaton remains unchanged, i.e.\ $\tau$ is holomorphic and completely specified by (\ref{tauBI}). On the other hand, the relation between $F_{\it 3}$ and $\tilde F_{\it 3}$ to be modified by the coupling between the three-form fluxes and the gaugino bilinear in  the  D-brane action. The precise form of this term is calculated in appendix \ref{app:fermaction} by an accurate dimensional reduction of the D-brane fermionic action \cite{Dbraneferm} (see also \cite{camara,dwsb} for previous studies)
\be
\label{localferm}
S^{\rm ferm}_{\rm D7}= -\frac{i\,\pi}{2\ell^4_s}\, \int_{X_4}\d^4 x\,\sqrt{-g_4}\,\langle \bar S\rangle \int_D (G_{\it 3}\cdot \Omega)\, J\wedge J \, +\, \text{c.c.}  +\ldots
\ee
The relevant terms in the action are
\be\label{efflagr}
\call^\prime=\frac{2\pi}{\ell^8_s}\, \int_M\Big\{-\frac12\, e^{\phi} * F_{\it 3}\wedge F_{\it 3}+\Re[{\ell^4_s}\langle \bar S\rangle \delta^{(0)}_D\,\Omega]\wedge F_{\it 3}+\tilde C_{\it 2}\wedge (\d F_{\it 3}+\ldots)\Big\}\ .
\ee
Here the first and the last terms are the usual action in the bulk and the second term comes from (\ref{localferm}).

The last term in (\ref{efflagr}) defines $\tilde{C}_{\it 2}$ as the Lagrange multiplier in front of the BI for the RR three-form $\d F_{\it 3}+\ldots=0$.
In this action we consider $\tilde C_{\it 2}$ and $F_{\it 3}$ as the independent  dynamical fields. Thus the equation of motion for  $\tilde C_2$
is the BI for $F_{\it 3}$ while the equation of motion for $F_{\it 3}$ provides the definition of $\tilde{F}_{\it 3}\equiv \d \tilde C_{\it 2}$ -- the relation we are looking for
\be\label{dualflux3}
\tilde F_{\it 3}\equiv\d \tilde C_{\it 2}=-e^\phi*F_{\it 3}+\Re[{\ell^4_s}\langle \bar S\rangle\delta^{(0)}_D\,\Omega]\, .
\ee
While the first term $-e^\phi*F_{\it 3}$  is coming from (\ref{defdualF}) the second localized term $\Re[{\ell^4_s}\langle \bar S\rangle\delta^{(0)}_D\,\Omega]$ is the extra local contribution we advertised above.

With help of (\ref{nablatheta}) this equation can be easily solved, yielding
\bea\label{FHrel}
F_{\it 3}&=&-e^{-\phi}*H+\frac12\Im\big[e^{-\phi}\,(\nabla^m\bar\theta_m)\,\Omega] = \\
\label{RR3flux} \nonumber
&=&- \frac14 e^{-\phi}\Im(\del\theta^m\wedge \iota_m\bar\Omega)-\frac14 e^{-\phi}\Im\big[(\partial_m\phi)\bar\theta^m\Omega\big]+\frac14\, e^{-\phi}\,  \Im[(\nabla^m\bar\theta_m)\Omega]\ .
\eea
Together with (\ref{defHflux}) this can be combined into the complex three-form
\be\label{uwIASDflux}
G_{\it 3}\equiv F_{\it 3}+ie^{-\phi}H=\frac{i}{4} e^{-\phi}\del\theta^m\wedge \iota_m\bar\Omega+\frac{i}{4} e^{-\phi}(\nabla_m\phi)\,\bar\theta^m\,\Omega+\frac{i}{4} e^{-\phi}(\nabla^m\theta_m)\,\bar\Omega\ .
\ee
Notice that the first two terms are IASD and have the same structure as the IASD field found in \cite{bau10}, while the last term  is ISD and completely localized on $D$.
Of course this is not a coincidence. The relation between our results and findings of \cite{bau10} will be explained in section \ref{sec:warpingD3}.

The equations of motion for $G_{\it 3}$ follows from the supersymmetry conditions and therefore we can be sure that they are satisfied (although we will check this in a moment). The biggest challenge is the BI
\be\label{deformedRR}
\d F_{\it 3}+H\wedge F_{\it 1}+ B\wedge \sum_{a} \delta^{\it 2}_{D_a} =0
\ee
which does not follow from the supersymmetry conditions. Since the beta-deformation completely fixes $G_3$  there are no degrees of freedom we can adjust to satisfy the Bianchi identity. Luckily
the identity  (\ref{deformedRR}) is satisfied by (\ref{uwIASDflux}). Unfortunately this is not the case already at the second order in $\beta$ as explained in appendix \ref{sec:allorder}.

As a last step we would like to check explicitly that the equations of motion are satisfied. The equation of motion for $F_{\it 3}$ is nothing but the condition $\d\tilde{F}_{\it 3}=0$
\be\label{modifiedRR}
\d (e^{\phi}*F_{\it 3})=\d\Re\big(\ell_s^4\langle\bar S\rangle \delta^0\,\Omega\big)\ .
\ee
The equation of motion for $B$ (\ref{HEoM}) does not follow from the supersymmetry conditions so easily.
Of course we can obtain it just by varying the action with respect to $B$.
Keep in mind that in addition to the conventional terms in the bulk the full action also includes the local term on the D7-branes
\be\label{effectlagr}
 -\frac{2\pi}{\ell^8_s}\,\sum_{a} \int_M e^{\phi}\,\Re T_{\it 4}\wedge \delta^{\it 2}_{D_a}\ .
\ee
This is nothing but the DBI action of the D7-branes. It depends on $B$ and therefore gives a localized contribution to (\ref{HEoM}).
Although $ e^{\phi}\,\Re T_{\it 4}$ (where index $\it 4$ means we a picking the four-form) is the value of the DBI action on shell, in fact,
varying $e^{\phi}\,\Re T_{\it 4}$ with respect to $B$  using (\ref{defps}) (which implies that $\delta T= \delta B\wedge T$)
would produce the same result as the conventional DBI action.
We do not need to add the CS term because in our formalism the equations for $F$ (\ref{RREoM}) and $B$ (\ref{HEoM}) do not have the terms proportional to the BI. As a result we have
\bea\label{deformedH}
\d(e^{-\phi}*H)-e^{\phi}*F_{\it 3}\wedge F_{\it 1}-B\wedge \sum_a \delta^{\it 2}_{D_a} +e^{-\phi}\d\Im (\ell_s^4 \langle\bar S\rangle \delta^{(0)}_D \Omega)=0\ .
\eea
This equation is different from (\ref{HEoM}) only by the last term which comes from the coupling of the gaugino bilinear to the three-form flux (\ref{localferm}).

To make the connection with the conventional supergravity EOM we combine  (\ref{modifiedRR},\ref{deformedH}) and  (\ref{deformedRR}) into
\bea
\label{EOM}
{1\over 2}\left[\d\Lambda+\partial\phi\wedge (\Lambda+\overline{\Lambda})\right]=e^{-\phi} \d\big(\ell_s^4\langle{S}\rangle\, \delta^{(0)}_D \,\bar\Omega \big)+2iB\wedge \sum_a \delta^{\it 2}_{D_a}\ ,\\
\Lambda=2G_-,\quad G_-\equiv(*G_3-iG_3)\ .
\eea
This equation would follow from the conventional supergravity action corrected by the local term (\ref{localferm}) and the DBI+CS action of the D7-branes. In the last term of (\ref{EOM}) we also assumed that $B$ as given by (\ref{Bfield}) is of $(2,0)+(0,2)$ type and  hence the contribution of the DBI and CS terms are equal to each other, since  $*_4B+B=2B$.

To check that (\ref{EOM}) is satisfied it is helpful to notice that the first term in the r.h.s. of (\ref{EOM})
is localized on the stack of the D7-branes where the gaugino condensate is taking place, while the second term is localized on all D7-branes present.
Indeed the first term will be balanced by the singularity of $\theta$, while the second term is balanced by the singularity of the dilaton.
This can be checked straightforwardly using (\ref{nablatheta}) and (\ref{tauBI}). Many useful details can also be found in \cite{bau10}.

Similarly, the BI (\ref{deformedRR}) can be rewritten as
\bea
\label{newBI}
\d G_{\it 3}+\partial\phi\wedge(G_{\it 3}-\overline G_{\it 3})+B\wedge \sum_{a}\delta^{\it 2}_{D_a}=0\ .
\eea
Using the formal similarity between (\ref{newBI}) and (\ref{EOM}) and between $G_{\it 3}$ and $\Lambda$,
to check (\ref{newBI}) one only has to show that the external derivative of the ISD part of $G_{\it 3}$ (the last term in (\ref{uwIASDflux})), taken assuming the dilaton is constant,
is equal to ${i\over 4}e^{-\phi}\d\Im (\ell_s^4 \langle\bar S\rangle \delta^{(0)}_D \Omega)$. This identity immediately follows from (\ref{nablatheta}).

Eventually we notice that the the D7-branes  remain supersymmetric in the new background  with no world-volume gauge field turned on, despite the presence of $(2,0)+(0,2)$ B-field in the bulk. To see this,
we use the supersymmetry conditions written in terms of the pure spinors $\calz,T$  \cite{lucal,lucasup}:
\be
[(\iota_X\calz)|_{D_a}]_{\it 4}=0\quad \forall X\in T_M\, ,\qquad  [\calz|_{D_a}]_{\it 3}=0\, ,\qquad [\Im T|_{D_a}]_{\it 4}=0\ .
\ee

What happens if we add the O7-branes into the picture? The BI for the axion-dilaton (\ref{tauBI}) will now take the form
\bea
\d F_{\it 1}= \left(\sum_{a} \delta^{\it 2}_{D_a}-4\sum_b \delta^{\it 2}_{O_b}\right)\ ,
\eea
and therefore the delta-function contributions coming from the singularity of dilaton in (\ref{deformedRR},\ref{deformedH},\ref{EOM},\ref{newBI})
will change accordingly. Similarly the localized terms $B\wedge \sum_a \delta^{\it 2}_{D_a}$ in the Bianchi identity (\ref{deformedRR},\ref{newBI}) will be modified
into
\bea
B\wedge \left(\sum_{a} \delta^{\it 2}_{D_a}-4\sum_b \delta^{\it 2}_{O_b}\right)\ .
\eea
Hence the BI still will be satisfied by (\ref{uwIASDflux}).
At the same time the localized terms $B\wedge \sum_a \delta^{\it 2}_{D_a}$ in the equation of motion (\ref{deformedH},\ref{EOM}) will remain the same. Naively this means that in the presence of the O7-planes the three-form flux $G_3$ (\ref{uwIASDflux})
will not solve the equations of motion due to the unbalanced terms of the form
\bea
\label{unbalanced}
B\wedge \sum_b \delta^{\it 2}_{O_b}\ .
\eea
In fact these terms will vanish because the $B$-field must be odd on the O7-plane and hence the pullback of $B$ on $O_b$ is zero.
The only tricky case here is when $B$ is singular on $O_b$. This can happen if the O7-plane is accompanied by some D7-branes with the gaugino condensate on it. Then the resulting
$B$, locally, is given by (\ref{Bfield}). Clearly such $B$ is odd under the reflection in the direction orthogonal to $D$. To regularize (\ref{unbalanced})
we need to move the D7-branes away from the O7-plane by a small distance $\epsilon$ and consider the two copies of the D7-brane stack  located on both sides of the O7-plane.
Now the relevant term from (\ref{unbalanced}) will look like $(B_++B_-)\wedge \delta^{\it 2}_{O}$ where $B_{\pm}=B(\pm \epsilon)$ is the value of $B$ on the two stacks of D7s.
Clearly $B_\pm=-B_\mp$ in the directions along $O$ and in the limit when $\epsilon$ goes to zero (\ref{unbalanced}) vanishes. Eventually we conclude that (\ref{uwIASDflux})
is universal and that it describes the linear deformation of the background when both the D7-branes and the O7-planes are present.


\subsection{The backreaction of condensing D7-branes: a summary}

This concludes our description of the supersymmetric background deformation induced by gaugino condensation on  D7-branes, up to first order in $\langle S\rangle$. In summary, the gaugino condensate has the effect of deforming the supersymmetry of the background in a way which corresponds to the generalized deformation of the classical complex structure. This is encoded in the non-vanishing $\calz_{\it 1}\neq 0$ in (\ref{newcalz}). This implies that the associated generalized complex structure $\calj: T_M\oplus T^*_M\rightarrow T_M\oplus T^*_M$ has the form
\be\label{betaGC}
\calj=\left(\begin{array}{cc}  -I & -(I\beta+\beta I^T)\\ 0 & I^T\end{array}\right)
\ee
where $I\equiv I^m{}_n$ is the ordinary complex structure of the undeformed space and $\beta=\beta^{2,0}+\beta^{0,2}$ is  specified by (\ref{deftheta}). $\calj$ is integrable everywhere, except for $D$, and it is a genuine generalized complex structure because of the non-vanishing off-diagonal term in (\ref{betaGC}).

Besides the deformation of the integrable structure, the presence of $\langle S\rangle$ gives rise to the  $G_{\it 3}$-flux (\ref{uwIASDflux}), which contains a bulk IASD component and a localized ISD component. On the other hand, metric, dilaton and warping are unmodified to first order.


\section{Condensing D7-branes and the D3-brane superpotential}
\label{sec:warpingD3}
The generalized complex structure associated with the deformation of $\calz_{\it 1}\neq 0$ in the presence of $\langle S\rangle\neq 0$ has a simple physical interpretation in the presence of D3-branes.
It is well known that D3-branes affect the non-perturbative superpotential $W_{\rm np}$ generated by condensing D7-branes \cite{ganor96,baumann0607} and therefore experience a force. Hence the D3-branes are not compatible with the supersymmetry of the background because of the non-vanishing F-terms. This can only happen when the manifold is not a classical CY but a generalized one. In our formalism the relation between the deformation of the complex structure $\calz_{\it 1}\equiv -\frac{\ell_s^4}{\pi}\,\d w$ and the force on the D3-branes is straightforward: $w$ is the superpotential of the theory on a probe D3-brane \cite{lucasup}. This same conclusion can be reached based by analyzing the force  on the probe D3-brane in the background with the IASD flux (\ref{uwIASDflux}) \cite{bau10}.

Since D3-branes source a non-trivial warping, in order to properly address this question we modify the discussion of the section (\ref{sec:firstord}) by introducing a non-trivial  warping into the story. Although the full analysis is beyond the scope of this paper, the main points are as follows.

First, we start with more general GKP backgrounds \cite{gkp}, which include a non-trivial warping $e^{4A_E}$ (index E stands for Euclidean frame), ISD  $G_{\it 3}$ and $F_{\it 5}$ fluxes. In this case the string frame metric has the form
\be
\d s^2=e^{\phi/2}\big( e^{2A_{\rm E}}\d x^\mu\d x_\mu+e^{-2A_{\rm E}}\d s^2_M\big)\ .
\ee
This is different from (\ref{Ewarpedmetric}) by the non-trivial factors $e^{2A_{\rm E}}$ .
The internal metric $\d s^2_M$ and the axion-dilaton are the same as in the unwarped case of section \ref{sec:D7back}. There is also a RR five-form flux $F_{\it 5}=*\d e^{-4A_{\rm E}}$. The pure spinor $\calz$ for this vacua still takes the form (\ref{psD7}).

If we now include the effect of the gaugino condensate on the D7-branes, repeating the argument that lead us to (\ref{D7def}), we still get the deformation
\bea
\d\calz_{\it 1}=-2i\,\ell_s^4\, \langle S\rangle\, \delta^{\it 2}_D\ .
\eea
The solution $\calz_{\it 1}=\theta$ is the same with $\theta$ given by (\ref{loctheta}) and (\ref{intw}). Hence the superpotential $w$ for the  D3-brane remains unmodified.
In \cite{bau10} this superpotential was found to be related to the IASD flux of the form (compare with (\ref{uwIASDflux}))
\bea
\label{Lambda}
\Lambda\equiv 2e^{4A_E}(*G_{\it 3}-iG_{\it 3})=e^{-\phi}\del\theta^m\wedge \iota_m\bar\Omega+ e^{-\phi}(\nabla_m\phi)\,\bar\theta^m\,\Omega\ .
\eea
Let us emphasize here that (\ref{Lambda}) does not specify $G_{\it 3}$ completely, but only its IASD part. The same combination $\Lambda$ is the only part of the 3-flux constrained by the equation of motion. A simple generalization of (\ref{EOM}) to the case with warping gives
\bea
\label{EOMwarping}
{1\over 2}\left[\d\Lambda+\partial\phi\wedge (\Lambda+\overline{\Lambda})\right]=e^{-\phi} \d\big(\ell_s^4\langle{S}\,\rangle \delta^{(0)}_D\,\bar{\Omega} \big)+{i\ell_s^8\over 2\pi}\sum_a {\delta S_{D7}\over \delta B}\wedge \delta^{\it 2}_{D_a}\ .
\eea
This equation is solved by (\ref{Lambda}) provided the local terms $\sum_a {\delta S_{D7}\over \delta B}\wedge \delta^{\it 2}_{D_a}$ are balanced by the appropriate behavior of the flux and the axion-dilaton at the locations of the D7-branes $D_a$. This relation together with the BI should fix the ISD part of $G_{\it 3}$.
Alternatively one can find it by solving the supersymmetry conditions (\ref{defsusycond2},\ref{defsusycond3}) similarly to the case of trivial warping $e^{4A_E}=1$ considered in this paper. To complete (\ref{Lambda}) to the full solution is an interesting  problem for the future.

The explicit expression for $w$ can be used to determine the dependence on the mobile D3-branes of the non-perturbative superpotential $W_{\rm np}$ generated by the condensing D7-branes. Given $w$, we can find $W_{\rm np}$ by matching the force on the D3-branes as originating from $w$ and $W_{\rm np}$.
Let us consider $K$ mobile D3-branes located at $\hat z_{k},\ k=1,\ldots,K$ in the internal manifold. These positions generically enter the D7-branes gauge coupling through the threshold corrections. We assume that the non-perturbative potential is governed by the gaugino condensate such that
\be\label{lowenD3}
W_{\rm np}(\hat z_{1},\ldots,\hat z_{K})={\mathcal N}\, \langle S\rangle\ ,
\ee
where ${\mathcal N}$ is some constant which depends on the details of the low-energy theory on the D7-branes and the condensate $\langle S\rangle$ depends on $\hat z_{k}$.
For the ${\rm SU}(N)$ theory ${\mathcal N}=N$. In general this coefficient depends on the beta-function of the low-energy field theory.

We use (\ref{lowenD3}) to find $ \langle S\rangle$ and $w$ as given by (\ref{intw}):
\be\label{intw2}
w(z;\hat z_{1},\ldots,\hat z_{K})={1\over \mathcal N}W_{\rm np}(\hat z_{1},\ldots,\hat z_{K})\,\log h(z)+w_0\ .
\ee
Here $w_0$ is some $z$-independent constant, which can depend on the locations of the mobile D3-branes $\hat{z}_k$.

We now impose the agreement for the force applied on the $k$-th D3-brane as calculated
using $w$ and $W_{\rm np}$
\bea\label{dersup}
\left.\frac{\del W_{\rm np}(\hat z_{1},\ldots,\hat z_{K})}{\del \hat z_{k}^i}=\frac{\del w(z;\hat z_{1},\ldots,\hat z_{K})}{\del z^i}\right|_{z=\hat z_{k}}\ .
\eea
These equations can be seen as a system of differential equations determining  $W_{\rm np}$. This can be easily integrated
\be\label{effsupred}
 W_{\rm np}(\hat z_{1},\ldots,\hat z_{K})=\cala\,\prod^K_{k=1}h^{1/\mathcal N}(\hat z_{k})\ ,
\ee
and $\cala$ is some $\hat{z}_k$-independent integration constant. $\cala$ can dependent on other chiral fields, e.g. the compactification  moduli.
In appendix \ref{sec:npsup} we revisit the derivation of $W_{\rm np}$ based on the logic of \cite{ganor96,baumann0607} and generalize it
to the case of arbitrary K\"ahler metric and holomorphic axion-dilaton. This result is in complete agreement with the expression (\ref{effsupred}).

The calculation above is based on the logic presented in \cite{tentofour,bau10}. Compared with \cite{tentofour}, here we clarify the distinction between $w$ and $W_{\rm np}$. We also extended the results of \cite{bau10} to backgrounds with holomorphic axion-dilaton.
More importantly, in \cite{bau10} the relation between $\partial w$ and $\partial W_{\rm np}$ was established only up to an overall coefficient. Here we confirmed this coefficient to be one, as in (\ref{dersup}). In particular this implies that the coupling between the three-form flux to the gaugino bilinear on the world-volume of the D7-branes (\ref{gauginoterm}) is exact,\ i.e. will not get loop corrections.

Clearly a similar  argument should  cover the non-abelian case, when  the D3-branes coincide, or other possible localized sources of the D3-brane charge like those originating from the D7 world-volume fluxes. The latter can play  a crucial role in the mechanism for generating Yukawa couplings on D7-branes through the non-perturbative effects \cite{npYuk}.


\section{Discussion}

In this paper we discussed how gaugino condensation on D-branes can be incorporated into a ten-dimensional picture.
We considered in detail the cases of gaugino condensation on  D5, D6 and D7-branes, wrapping some supersymmetric cycles in the internal six dimensional manifold.
Our analysis was local, in that we took the internal manifold to be non-compact. This corresponds to the $M_{\it P}\rightarrow \infty$ limit from the four-dimensional
point of view.

The discussion of D5-branes is focused on the description of the dynamical effect of the gaugino condensate $\langle S\rangle$ on the bulk complex structure. It was shown that  $\langle S\rangle\neq 0$ acts as a `source' for the complex structure, which is then dynamically deformed away from its tree-level value. In the mirror symmetric setting of D6-branes, the condensate  $\langle S\rangle\neq 0$ acts as a `source' for the symplectic structure. These results, based on local ten-dimensional arguments, have been compared with well-known results on geometric transition in the context of the gauge/gravity correspondence \cite{ks,mn,vafa,vafa2,martellimalda}.

The case of D7-branes displays some distinguished features. The ordinary complex structure of the tree-level background
is deformed into a  generalized complex structure. Moreover, it allows for a relatively simple perturbative analysis of this effect, with all ten-dimensional supersymmetry conditions under control, at least to first order. In particular, we have discussed how the supersymmetry conditions are modified by localized terms on the D7-branes.
Furthermore, we have shown that the deformed supersymmetry conditions are perfectly consistent with the equations of motion and Bianchi identity.
Our analysis bridges the gap between the approach of  \cite{tentofour}, which was based on supersymmetry, and the approach of \cite{bau10}, which was based on the supergravity equations of motion. In particular we confirm the prediction of \cite{bau10} that the gaugino condensate sources IASD $G_{\it 3}$ flux in the bulk, working in the more general framework with a holomorphic axion-dilaton.

There are a number of interesting open questions. For instance one should investigate possible SUSY breaking associated with the compactification effects. We touched this topic in section \ref{sec:susyb} where we argued that the SUSY breaking could have a topological origin. This issue may be related to the problem of extending the local solution to a global one. In the case of gaugino condensate on D7-branes, there is an additional complication. The first order solution described in section \ref{sec:firstord} is valid only in the neighborhood of a local patch of the divisor wrapped by the D7-branes supporting the gaugino condensate. Even in a purely local approach with non-compact internal space,  extending our result to a complete solution may present complications similar to the topological obstructions affecting settings with D7-branes at the classical level \cite{cacha01,franco10}. The two problems appear similar and their solution could require a more detailed understanding of the D-brane setting and of the associated low-energy effective theory which could include other light modes besides the pure SYM sector. This understanding seems necessary to have a picture of the large $N$ geometric transition  in the D7-brane setting.

Another important direction is to find the local description of the geometry around the D7-branes beyond the linear order in the gaugino condensate. Encouraged by the simple form of the linear deformation, we investigated in appendix \ref{sec:allorder} if a straightforward generalization of the same ansatz can solve the equations to higher orders in $\langle S\rangle$. Although all the supersymmetry conditions can be solved, the RR Bianchi identities
are solved only up to a term quadratic in $\langle S\rangle$.  {\em Formally} this can be interpreted as a delocalization of the D7-branes. To support this interpretation or to find the solution that solves the unmodified Bianchi identities would be a significant step forward.
In a recent paper \cite{heid10}, the same problem was addressed in the particular  setting
of four D7-branes and one O7-plane wrapping the $\mathbb{CP}^2$ in the CY complex cone over $\mathbb{CP}^2$. The non-perturbative in  $\langle S\rangle$ solution found therein
develops a singularity at finite radius, before reaching the putative location of the divisor with the condensing branes.
It would be interesting to understand the relation between the results presented there and here.

We leave a better understanding of these and other important issues to the future.



\vspace{3cm}

\centerline{\large\em Acknowledgments}

\vspace{0.5cm}

\noindent We thank M.~Bianchi, S.~Gukov, J.~Heckman, B.~Heidenreich, D.~Jafferis, A.~Lionetto, J.~Maldacena, F.~Marchesano, L.~McAllister, D.~Sorokin, A.~Tomasiello, C.~Vafa and A.~Zaffaroni for useful discussions and correspondence and B. Heidenrei for reading the manuscript. A.D.~thanks the theory group at Ludwig Maximilian Universit\"at for hospitality while this work was initiated. L.M.~would like to thank the Physics Department of Universit\`a di Parma for kind hospitality during the course of this work.
A.D.~gratefully acknowledges  support from the Monell Foundation, the DOE grant DE-FG02-90ER40542, and the Ministry of Education and Science of the Russian Federation under contract 14.740.11.0081.  The work of L.M.~was partly supported by the ERC Advanced Grant n.226455 ``Superfields'', by the Italian MIUR-PRIN contract 2007-5ATT78 ``Symmetries of the Universe and of the Fundamental Interactions'' and by the Cluster of Excellence ``Origin and Structure of  the Universe'' in M\"unchen, Germany.

\vspace{3cm}

\newpage


\begin{appendix}

\section{D7-brane fermions and $3$-form flux}
\label{app:fermaction}

The aim of this appendix is to compute the relevant terms describing the coupling of the D7-brane fermionic bilinear
to the background three-form flux.

The fermionic sector of the $\kappa$-symmetric D-brane action, at the quadratic level in fermions, was computed in \cite{Dbraneferm}.
In the case of  D7-branes wrapping a divisor $\Sigma_8\subset X_{10}$, with the gauge-invariant worldvolume flux $\calf=B|_{\Sigma_8}+2\pi\alpha^\prime\,{\rm F}$, it is
\be\label{faction}
S^{\rm ferm.}_{\rm D7}\,=\, \frac{i\pi}{\ell_s^8}\int_{\Sigma_8}\d^8\sigma\, e^{-\phi}\sqrt{\det(g_{10}|_D+\calf)}\,\bar\theta[1-\Gamma(\calf)]\left(\calm^{\alpha\beta}\Gamma_\alpha \cald_\beta-\frac12\calo\right)\theta
\ee
where $\alpha,\beta,\ldots$ are world-volume indices on  $\Sigma_8$. $\cald_\alpha$ and $\calo$ are the pullback of operators acting on bulk fermions, whose explicit form can be found in Appendix A of \cite{dwsb}.
The doublet
\bea
\theta=\left(
\begin{array}{c}\theta_1 \\
\theta_2\end{array}\right)
\eea
 is the GS-spinor on the D7-brane, while $\calm^{\alpha\beta}$ denotes the inverse of $\calm:= g_{10}|_{\Sigma_8}+\sigma_3\calf$. Eventually
$\Gamma(\calf)$ is the $\kappa$-symmetry operator, which for the D7-brane takes the form:
 \be
 \Gamma(\calf)=\sum_{q+r=4}\frac{(-)^{r+1}({\rm i}\sigma_2)(\sigma_3)^r\epsilon^{\alpha_1\ldots\alpha_{2q}\beta_{1}\ldots\beta_{2r}}}{q!(2r)!2^q\sqrt{-\det(g+\calf)}}\calf_{\alpha_1\alpha_2}\cdots \calf_{\alpha_{2q-1}\alpha_{2q}}\Gamma_{\beta_{1}\ldots\beta_{2r}}\ . \label{chiralB1}
 \ee

In what follows we start by assuming the background to be a general supersymmetric IIB background of the form discussed in section \ref{sec:genback}. The D7-branes fill the four-dimensional space $X_4$ and wrap the  four-cycle $D\subset M$. In order to analyze  the fermion bilinear, we employ the following $\kappa$-fixing gauge
\be\label{kfix}
\bar\theta\Gamma(\calf)\,=\,-\bar\theta\ .
\ee
Then, the fermionic action takes the form
 \be\label{faction2}
S^{\rm ferm.}_{\rm D7}\,=\, \frac{2\pi i}{\ell_s^8}\,\int_{X_4}\d^4 x \sqrt{-g_4}\int_D \d^4\sigma\, e^{4A-\phi}\sqrt{\det(g|_\Sigma+\calf)}\,\bar\theta\big(\Gamma^\mu \cald_\mu+\calm^{\alpha\beta}\Gamma_\alpha \cald_\beta-\frac12\calo\big)\theta
\ee

To extract the dependence on the four-dimensional fermions we need to known the Kaluza-Klein reduction form of  $\theta$. It is built with the help of the  supersymmetry generators $\epsilon=(\epsilon_1,\epsilon_2)$ (here $\zeta$ is a constant four-dimensional chiral spinor)
\be
\epsilon_1=\zeta\otimes \eta_1+\text{c.c.}\, ,\quad \epsilon_2=\zeta\otimes \eta_2+\text{c.c.}
\ee
, specified by the chiral six-dimensional spinors $\eta_1$ and $\eta_2$ as follows.
 Supersymmetry requires the D-brane to satisfy the $\kappa$-symmetry condition $\bar\epsilon\Gamma(\calf)=\bar\epsilon$.
 Using this and the $\kappa$-fixing (\ref{kfix}), we are led to identify the four-dimensional gaugino $\lambda_{\rm D}$ (the Dirac spinor) with the four-dimensional part of  $\theta$
\be\label{gino}
\theta_1\,=\,\frac{\ell_s^2}{4\pi}\, e^{-2A}\lambda_{\rm D}\otimes \eta_1 +\ \text{c.c.}\quad\quad\quad \theta_2\,=\,-\frac{\ell_s^2}{4\pi}\,e^{-2A}\lambda_{\rm D}\otimes\eta_2+\ \text{c.c.}
\ee
As a check, one can apply the supersymmetry transformations on the D7-brane which was found in \cite{Dbraneferm} to get  the standard four-dimensional supersymmetry transformations relating the four-dimensional gauge field to $\lambda_D$. In order to fix normalization in (\ref{gino}) we calculate the   kinetic term which reads
\be
\frac{i}{2\pi}\,\Re\alpha\, \bar\lambda_{\rm D}\gamma^\mu\partial_\mu\lambda_{\rm D}\ .
\ee
Here $\gamma^\mu$ are the four-dimensional Dirac matrices associated with the flat metric $\d x^\mu\d x_\mu$ and $\alpha\equiv\alpha(\calt)$ is the holomorphic gauge coupling (\ref{holgc}).

The gaugino $\lambda_D$ is chiral in four dimensions (here we choose a particular representation of the gamma matrix algebra such that $\gamma^5$ is diagonal)
\be
\lambda_{\rm D}=\left(\begin{array}{c} 0 \\ \bar\lambda^{\dot\alpha}\end{array}\right)\, .
\ee
To facilitate the comparison with the literature we rewrite the result for the kinetic term in the Weyl representation
\be
-\frac{i}{2\pi}\,\Re\alpha\, \lambda\sigma^\mu\partial_\mu\bar\lambda\ .
\ee
It has the canonical normalization, as in (\ref{SYMexp}).

We now compute the coupling  of $S={\lambda\lambda\over 16\pi^2}$ to the three-form flux $G_3$. Since $\bar\theta\theta\sim \langle S\rangle $, in (\ref{faction2}) we can  keep classical configurations for the background metric, dilaton and the D7-brane embedding. We also set $\calf=0$ for simplicity.  After some algebra we arrive at the following term
\bea\label{gauginoterm}
S_{\rm D7}&\supset&\frac{1}{32\pi\ell_s^4}\int_{X_4}\d^4 x \,\Big(\lambda_{\rm D}^TC\lambda_{\rm D}\,\int_D G_{\it 3}\cdot \Omega\, J\wedge J \, +\, \text{c.c.}  \Big) \cr
&=& -\frac{i}{32\pi\ell_s^4}\int_{X_4}\d^4 x \,\bar\lambda_{\dot\alpha}\bar\lambda^{\dot\alpha}\,\int_D G_{\it 3}\cdot \Omega\, J\wedge J \ +\ \text{c.c.}
\eea
where $\bar\lambda_{\dot\alpha}\bar\lambda^{\dot\alpha}=\varepsilon_{\dot\alpha\dot\beta}\,\bar\lambda^{\dot\beta}\bar\lambda^{\dot\alpha}$.


\section{An all-order deformation with smeared D7-branes}
\label{sec:allorder}

We now comment on the possibility of extending  the first-order supersymmetric
solution found in section \ref{sec:firstord} to higher orders. The most natural strategy, which we want to explore, is to simply `exponentiate' the first order $\beta$-deformation introduced therein.
The problem with this strategy is that the linear solution in section  \ref{sec:firstord} heavily relied on peculiarities of the classical backgrounds of section \ref{sec:D7back}. Hence the correct second order deformation may be more complicated than the naive exponentiention  of the linear one.

Nevertheless it makes sense to investigate what happens if we apply the beta deformation beyond the linear order.
The resulting finite $\beta$-deformation on $T_M\oplus T_M^*$ spinors (i.e.\ polyforms, like $\calz$ and $T$) is  the operator
\be
e^{\beta}=1+\iota_\beta+\frac12\iota_\beta\iota_\beta+\frac1{3!}\iota_\beta\iota_\beta\iota_\beta\ ,
\ee
with the new spinors $\calz$,$T$ of the form
\be
\label{beta-deformation}
\calz=e^{\beta}\cdot \calz^0\, ,\qquad T=e^{\beta}\cdot T^0\ .
\ee
Here we labeled the undeformed pure spinors, provided by (\ref{psD7}), with $0$.
The transformation (\ref{beta-deformation}) has a natural advantage of keeping $\calz$ and $T$ pure and compatible at all orders in $\beta$.

Since $\beta$ has only $(2,0)$ and $(0,2)$ components, the action on $\calz^0\equiv \Omega$ stops at first order. Therefore the deformed $\calz$ coincides with (\ref{betacalz}) even for finite $\beta$ and, by choosing $\theta=\beta\lrcorner \Omega$ as in (\ref{thetadef})-(\ref{intw}), equation (\ref{D7def}) is also satisfied for finite $\beta$.

Let us now turn to the condition (\ref{defsusycond2}). The warping $e^{A}$ can be extracted from $\calz$ and $T$ by the following formula
\be
e^{6A}=\frac{\langle \calz,\bar\calz\rangle}{\langle T,\bar T\rangle}\ .
\ee
Because $e^{\beta}\cdot$ preserves the Mukai pairing the warping is not changed and we have
\be
e^{2A}=e^{\phi_0/2}\ .
\ee
Hence, $e^{2A}\Im T$ takes the form
\bea\label{imt2}
e^{2 A}\Im T&=&e^{\phi_0/2}\Im T^0-\frac{1}{3!}\,e^{\phi_0}(\iota_\beta+\frac12\iota_\beta\iota_\beta)(J\wedge J\wedge J)\cr
&=& e^{\phi_0/2}\Im T^0+\frac{i}{8}\, (\iota_\beta+\frac12\iota_\beta\iota_\beta)(\Omega\wedge \bar\Omega)\cr
&=&e^{\phi_0/2}\Im T^0+\frac{1}{4}\, \Im(\bar\theta \wedge \Omega)+\frac{i}{8}\, \theta\wedge\bar\theta\ .
\eea
By taking the exterior  derivative we then find
\be\label{Dflatall}
\d(e^{2 A}\Im T)=\frac14\, \ell_s^4\,\delta^{\it 2}_D\wedge\big[ \langle S\rangle\,\bar\theta + \langle \bar S\rangle\,\theta \big]\ .
\ee
In order to preserve supersymmetry, we must impose $\d(e^{2 A}\Im T)=0$. In fact, by using the local solution (\ref{loctheta})-(\ref{intw}), the singular three-form in the right-hand-side of (\ref{Dflatall}) is zero. To understand this we can multiply it by some probe three form $\chi$ and integrate over $M$. The result will be an integral of the form $<\bar S>\int_D \theta \wedge \chi +c.c.$. Now, the one-form $\theta$ is proportional to $\d h/h$, where $h=0$ on $D$. Therefore the pull-back of $\theta$ on $D$ vanishes and (\ref{defsusycond2}) is satisfied for finite $\beta$ too!

We have seen that, already at first order in $\beta$, the tree-level supersymmetry conditions are modified by the localized terms on $D$. In section \ref{sec:firstord}, we have explicitly shown how to keep them under control to first order. However, understanding local terms comes after one understands how to satisfy the equations of motion in the bulk. In the rest of this section we simply neglect  all terms localized on $D$. In this approximation the tree-level supersymmetry conditions are expected to hold unmodified.

Now, we are left with the supersymmetry condition (\ref{defsusycond3}) that, neglecting localized terms induced by the gaugino condensate, is equivalent to the tree-level condition (\ref{susycond3}). As in section \ref{sec:firstord}, we can consider this condition as the definition of the supersymmetric RR fluxes, which in the case at hand are given by
\be\label{fluxdef2}
F=-\hat*\, \sigma\big[e^{-B} \wedge e^{-\phi_0}\d(e^{\phi_0}\Re T)\big]\ .
\ee
Remember that (\ref{fluxdef2})  implies that the equations of motion for $F$ away from $ D$ are automatically satisfied. However, the Bianchi identity (\ref{RRbi}) which we prefer to rewrite as
\bea
\d F^{\rm tw}=\d(e^B \wedge F)=-j
\eea
 needs to be separately verified.

In order to proceed (\ref{fluxdef2}), we would need the explicit form of the internal metric $\hat g$ (which enters $\hat *$) and of the $B$-field, which are encoded in $\calz$ and $T$. We will not need the new value of the dilaton. The beta-deformation is in fact a particular case of a more generic $O(6,6)$ transformation of the spinors $\calz,T$.
The general formula for the new $\hat{g}$ and $B$, valid for any $O(6,6)$ transformation is as follows. The $O(6,6)$ transformation can be represented by a matrix
\be
\calo=\left(\begin{array}{cc} a & c \\ b & d
\end{array}\right)
\ee
acting on  the generalized tangent bundle $T_M\oplus T_M^*$. The matrices $a,b,c$ and $d$ satisfy the $O(6,6)$ restrictions
\be\label{onnrestr}
a^Tb+b^Ta=0 \, , \qquad c^Td+d^T c=0 \, \qquad a^Td+b^Tc=\bbone \ .
\ee
By defining $E=\hat g+B$, the $O(6,6)$ action on $E$ is given by
\be\label{metricB-T}
E^0\quad\rightarrow\quad E=(b+d E^0)(a+cE^0)^{-1}\ .
\ee
Notice that this action (\ref{metricB-T}) is formally identical to the T-duality transformation. Moreover the transformation of the dilaton
preserves this formal analogy. In fact if $\calo$ is constant and has indexes only along the $U(1)$ invariant directions this transformation reduces to a continuous T-duality.
However, it should be clear that our case is different from the T-duality simply because
the elements of $\calo$ are generically not constant and there are no $U(1)$ isometries.

Now, the $\beta$-deformation introduced above can be seen as a O(6,6) transformation of the form
\be
\calo_\beta=\left(\begin{array}{cc} \bbone & -\beta \\ 0 & \bbone
\end{array}\right)
\ee
i.e.\ with $a=\bbone, b=0, c=-\beta, d=\bbone$.  See e.g. \cite{Butti:2007aq} for an application of this formalism to the study of the deformed SCFT's in AdS/CFT, along the lines of \cite{Lunin:2005jy}, when the constant $\beta$ defines a continuous T-duality.

By applying (\ref{metricB-T}) in the case of a $\beta$-deformation, we obtain
\be\label{defE}
E=E^0\cdot(\bbone-\beta E^0)^{-1}\simeq E^0+E^0\beta E^0 +\ldots
\ee
Up to quadratic order in $\beta$ the B-field is given by (\ref{Bfield}) while
\bea
\hat g_{i\bar \jmath}=e^{\phi_0\over 2}\Big[g^0_{i\bar \jmath}-{1\over 8}\left(g^0_{i\bar \jmath}|\theta|^2-\theta_i\bar\theta_{\bar \jmath}\right)\Big] \ .
\eea

To  calculate (\ref{fluxdef2}) we use somewhat sophisticated generalized geometry techniques.
The condition (\ref{D7def}) implies that  $\calz$ defines an integrable  generalized complex structure $\calj$ on $M\setminus D$. In general, it is possible to show that this is equivalent to the existence of a split of the exterior derivatives acting on polyforms into the sum of two generalized Dolbeault operators $\d=\del^\calj+\bar\del^\calj$ \cite{gualtieri}. In the case of interest
 when $\calj_\beta$ is associated with the holomorphic $\beta$-deformation of an ordinary complex structure, the generalized Dolbeault $\del^\beta\equiv \del^{\calj_\beta}$ operator is
\be\label{decbeta}
\del^\beta=\del-[\del,\iota_{\beta^{2,0}}]+[\bar\del,\iota_{\beta^{0,2}}]
\ee
where $\del$ is the ordinary Dolbeault operator associated with the  original complex structure.

By using (\ref{decbeta}), it is possible to rewrite the twisted field $F^{\rm tw}$  as $F^{\rm tw}=F^{\rm tw}_++F^{\rm tw}_-$ with  (see \cite{Tomasiello:2007zq} for details)
\be
F^{\rm tw}_+=i\del^\beta \Re T=i\del(e^{-\phi}-\frac12 J\wedge J)-\frac{i}{2}\delbar\big[\iota_{\beta^{0,2}}(J\wedge J)\big]-\frac{i}{2}\delbar\big[\iota_{\beta^{0,2}}\iota_{\beta^{2,0}}(J\wedge J)\big]
\ee
and $F^{\rm tw}_-=(F^{\rm tw}_+)^*$. We can now compute  $\d F^{\rm tw}$ by using the identity $\d F^{\rm tw}= \del^\beta F^{\rm tw}_-+ \delbar^\beta F^{\rm tw}_+ $.  After some work we get
\be\label{fincond}
\d F^{\rm tw}= -j+ \frac18\, \d\d^{\rm c}(e^{-\phi}g^{mn}\bar\theta_m\theta_n)
\ee
where $\d^c=i(\del-\delbar)$ and we have omitted the terms localized on $D$. This is an exact result valid to all orders in $\beta$.
The last term on the r.h.s.\  of (\ref{fincond}) should be vanishing in an ordinary supergravity background, a condition which does not appear to be satisfied. Hence, we see that a finite $\beta$-deformation fails to produce a ordinary supergravity background already at the second order. Yet the expression (\ref{fincond}) contains some nontrivial cancelations. In general one can expect $\d F^{\rm tw}$ to have the six- and four-form contribution, yet they vanish. The only contribution that remains is the two-form like the original $j$.
This suggests that we can formally promote the localized source (\ref{jd7}) to a partly smeared D7-brane current
\be
\hat\jmath=j-\frac18\, \d\d^{\rm c}(e^{-\phi}g^{mn}\bar\theta_m\theta_n)\ .
\ee
This would solve the BI, at least ignoring terms localized on the condensing brane.
The possible physical interpretation of this observation is not clear to us. We postpone the issues related to higher orders in $\beta$ for future investigations.


\section{Non-perturbative effects  and D3-brane superpotential}
\label{sec:npsup}

It is well known that D3-branes filling the space-time $X_4$ of a classical GKP background \cite{gkp} have flat classical potential and can move freely in the internal space. In the presence of Euclidean D3-branes (E3-branes, for short) or D7-branes this can be changed by non-perturbative effects. A non-trivial superpotential $W_{\rm np}$ for the D3-branes is generated and the D3-branes experience force.
The structure of the D3-brane location dependence of $W_{\rm np}$ generated on one E3 in the context of the F-theory compactifications without flux was first proposed in \cite{ganor96} by use of a monodromy argument. Later this result was  extended  in \cite{baumann0607} for gaugino condensation on D7-branes for some specific examples of warped conic backgrounds with trivial dilaton through an elaborate straightforward calculation. The logic used there was somewhat opposite to the one followed in section \ref{sec:warpingD3}. While in section \ref{sec:warpingD3} we studied the force on  D3-branes in the background including the backreaction by the D7-branes, in \cite{baumann0607} the D3-branes were treated as the source affecting the physics on the D7-branes.
In this appendix we would like to revisit the derivation of $W_{\rm np}$ using the same point of view as in \cite{baumann0607}. We provide an elegant derivation of (\ref{effsupred}) valid for any GKP background and for any divisor with a general $(1,1)$ primitive world-volume flux. Our derivation will also make a connection with the approach of \cite{ganor96}.

In what follows we do not assume the supersymmetry conditions (\ref{susycond}) to be satisfied. Hence the resulting $W_{\rm np}$ will be  equally valid in the presence of a mildly SUSY-breaking $G^{0,3}$ term that generates a non-vanishing expectation value of the  GVW superpotential. The latter is a crucial ingredient of the scenarios proposed in \cite{KKLMMT,LARGE} and like.  Furthermore, notice that the technical steps presented below can be straightforwardly applied to the  intrinsic F-theory settings, by using the dual M-theory picture with the non-perturbative superpotential generated by the M5-branes \cite{wittensup} in flux backgrounds of the kind described in \cite{beckerflux}.

\subsection{Supergravity derivation}
\label{sec:sugrader}

Let us consider a stack of D7-branes wrapping a divisor $D$ that undergo gaugino condensation.
 The non-perturbative superpotential generated by the gaugino condensation is given by
\be
\label{npcc}
W_{\rm np}=\Lambda^3=\mu_0^3\, {\rm exp}\left(-2\pi\alpha/{\mathcal N}\right)
\ee
where $\alpha$ is the field theory coupling constant (\ref{SYMcoupling}) defined at the  UV scale $\mu_0$ and $\mathcal N$ is some field theory dependent coefficient. In the case of pure $SU(N)$ SYM ${\mathcal N}=N$.
Furthermore, to describe the $W_{\rm np}$ generated by the E3-instanton it is enough to take  ${\mathcal N}=1$.

The gauge coupling $\alpha$ (\ref{holgc}) can be obtained by expanding the DBI+CS action for the D7-branes:
\be \label{realE3action}
\alpha= -\frac1{2\ell_s^4}\int_D \Big(e^{-4A_{\rm E}} J\wedge J-e^{-\phi}\calf\wedge \calf\Big)-\frac{i}{\ell_s^4}\int_D {C}\wedge e^{\calf} \ .
\ee
Here, we have used the form of $\calt=e^{-\phi}\exp(i\, e^{-2A_{\rm E}+\phi/2}J+B)$ for the GKP background with warping and  $\calf\equiv B|_D+2\pi\alpha^\prime {\rm F}$ is (1,1).

Consider now the case of a mobile D3-brane located at $y=\hat y$. Following the original idea of  \cite{Giddings:2005ff}, we aim to obtain (\ref{effsupred}) by extracting the dependence of $\alpha$ on the D3-brane coordinate $\hat y$. In other words, we need to compute how $\alpha$ changes under a small displacement of the mobile D3-brane. Then, using (\ref{npcc}) we will find how
$W_{\rm np}$ depends on $\hat{y}$. The D3-brane is mutually supersymmetric with the background before the non-perturbative effects on the D7-branes are taken into account.
The backreacted background remains of the GKP type. In fact the only part of the geometry affected by the D3-branes is the warp factor $A_{\rm E}$ which is determined by the following equation
\be\label{warpedlapl}
\nabla^2 e^{-4A_E}=\frac12\, e^\phi\, |G_{\it 3}|^2+*\,Q^{\rm loc}_{\rm D3}\ .
\ee
Here $ |G_{\it 3}|^2=\frac1{3!}G_{mnp}\bar G^{mnp}$ and
\bea
Q^{\rm loc}_{\rm D3}=(e^{-B}\wedge j)_{\it 6}&=&\ell_s^4\Big(\sum_{p\in \text{D3's}}\delta^{\it 6}_{y_{(p)}}-\frac14\sum_{q\in\text{O3's}}\delta^{6}_{y_{(q)}}\Big)+\frac12\sum_{a\in \text{D7's}}\delta^{\it 2}_{D_a}\wedge \calf_a\wedge \calf_a\ .
\eea
is the D3-brane localized charge\footnote{We work on the orientifold covering space. Furthermore, for simplicity, we omit the curvature contributions to the D3-brane charge induced on D7-branes and O7-planes.}
Neither metric nor axion-dilaton or three-form fluxes get affected when the D3 is moved. Hence, from (\ref{realE3action}) we get
\be\label{step1}
\frac{\delta\log |W_{\rm np}|^2}{\delta \hat y^m}=-{2\pi\over \caln \ell_s^4}\frac{\delta}{\delta \hat y^m}\int_D\, e^{-4A_E} J\wedge J\ .
\ee
In this section we denote the derivative with respect to the mobile D3-brane position with $\frac{\delta}{\delta \hat y}$, in order to better distinguish it from the derivatives with respect to the ordinary internal space coordinates. We can now rewrite the r.h.s.\ of (\ref{step1})
as an integral over the complete internal space $M$
\be\label{step2}
-{2\pi\over \caln \ell_s^4}\frac{\delta}{\delta \hat y^m}\int_M\, e^{-4A_E} J\wedge J\wedge \delta^{\it 2}_D={4\pi\over \caln \ell_s^4}\int_M\, \frac{\delta e^{-4A_E}}{\delta \hat y^m} {J\wedge J\wedge J\over 3!} \delta^{(0)}_D\ .
\ee
At the next step we vary the equation (\ref{warpedlapl}) with respect to $\hat y$
\be\label{sourced3}
\nabla^2 \frac{\delta e^{-4A_E}}{\delta \hat y^m}=\ell_s^4\,\frac{\delta}{\delta \hat y^m}\delta^{(0)}_{\hat y}\ ,
\ee
where
\bea
\delta^{(0)}_{y}\equiv *\delta^{\it 6}_{y}\ .
\eea
So far our calculation was not different from \cite{baumann0607}. The approach of \cite{baumann0607} was to solve the equation (\ref{sourced3}) explicitly expressing $\delta e^{-4A_E}$ through the Green's function on $M$. Then the corresponding result was substituted into (\ref{step1}) and integrated over $D$. Although conceptually simple, these steps require the metric on $M$ to be sufficiently simple to admit an explicit expression for the Green's function. In fact there is a more elegant way to proceed which is equally good for any K\"ahler metric on $M$. The idea is to use the Poicar\'e-Lelong equation mentioned in section \ref{sec:firstord} to express the delta-function $\delta^{(0)}_D$ localized on $D$ through the holomorphic section $h(z)$ of the line bundle that defines $D$ through $h=0$
\be\label{LH}
\delta^{(0)}_D=\frac{1}{4\pi}\nabla^2 \log |h|^2\ .
\ee
Naively this equation depends on the metric, but in fact this dependence is a phantom. This relation follows from $\bar\partial \partial  \log h=2\pi i \delta^{\it 2}_D$
which is metric independent.
After substituting (\ref{LH}) into (\ref{step2}), integrating by parts and using (\ref{sourced3}), we obtain
\be\label{step3}
\frac{\delta\log |W_{\rm np}|^2}{\delta \hat y^m}=\frac{1}{\caln}\frac{\delta}{\delta \hat y^m}\int_M {J\wedge J\wedge J\over 3!}\log|h|^2 \delta^{(0)}_{\hat y}=\frac{1}{\caln}(\del_m\log |h|^2)|_{y={\hat y} }\,\ .
\ee

This, together with the assumption that the superpotential $W_{\rm np}$ is holomorphic, is enough to prove that $W_{\rm np}=\cala\, h^{1/\caln}(\hat z)$ which matches (\ref{effsupred}). The extension to the case of more than one D3-brane is straightforward. The resulting relation for the $k$-th D3-brane
\be\label{step4}
\frac{\delta\log |W_{\rm np}|^2}{\delta \hat y^m_k}= \frac1\caln\,(\del_m\log |h|^2)|_{y={\hat y}_k }\ ,
\ee
coincides with (\ref{step3}). Clearly the superpotential is given by (\ref{effsupred}).

\subsection{Imaginary part and holomorphicity of $W_{\rm np}$}

So far we merely assumed holomorphicity of $W_{\rm np}$ based on its four dimensional interpretation.
In fact this too can be proved if one considers the imaginary part of $\log\, W_{\rm np}$ similarly to the calculation of the real part above. Indeed, we will presently show that the variation of the superpotential with respect to the antiholomorphic coordinate $\hat{\bar z}^{\bar\imath}$ vanishes
\be\label{holsup}
\frac{\delta W_{\rm np}}{\delta \hat{\bar z}^{\bar\imath}}=0\ .
\ee
This would be sufficient to conclude that (\ref{effsupred}) is the only solution of (\ref{step4}), obtaining the desired result. In what follows we consider only one D3-brane, although the generalization to many D3-branes is straightforward.

Let us start with the purely imaginary part of the non-perturbative superpotential as given by (\ref{npcc}) and (\ref{realE3action})
\be\label{imE3action}
\Im\log W_{\rm np}= \frac{2\pi}{\ell_s^4}\int_D \Big(C_{\it 4}+C_{\it 2}\wedge \calf+\frac12C_{\it 0}\, \calf\wedge \calf\Big)\ .
\ee
The treatment of $\Im\, \log W_{\rm np}$ requires some care, since the RR potentials are not globally defined. In particular,
the  D3-brane acts a monopole-like source in the BI
\be\label{F5bianchi}
\d F_{\it 5}+H \wedge F_{\it 3}=Q^{\rm loc}_{\rm D3}\ ,
\ee
 and this is exactly the reason why $C_{\it 4}$ and hence (\ref{imE3action}) depend on the D3-brane position. This is directly related to the approach of \cite{ganor96} which was based on the observations that $\Im\log W_{\rm np}$ should acquire a non-trivial shift by $2\pi$ when the D3-brane encircles around the D7-branes.

Our starting point is the BI (\ref{F5bianchi})
\be
\frac{\delta \d F_{\it 5}}{\delta \hat y^m}\,=\,-\ell_s^4\d(\iota_m\delta^{\it 6}_{\hat y})\ .
\ee
We identify $dC_{\it 4}$ with the closed part of $F_{\it5}$. Then the variation of $dC_{\it 4}$ with respect to a small displacement of the D3-brane is
\be\label{realinfBI}
\frac{\delta \d C_{\it 4}}{\delta \hat y^m}\,=\,\frac{\delta F_{\it 5}}{\delta \hat y^m}+\ell_s^4\iota_m\delta^{\it 6}_{\hat y}\ .
\ee
In particular we can take $ \delta C_{\it 4}$ to be of the $(2,2)$ type.
This is because the possible $(3,1)$ and $(1,3)$ contribution can be get rid of with help of the appropriate gauge transformation.
Indeed the possible $\delta C^{1,3}$ would obviously be $\delbar$-closed and it would then contribute to (\ref{realinfBI}) through the $\del$-exact and $\delbar$-closed piece $\d\delta C^{1,3}=\del\delta C^{1,3}$. With help of the $\del\delbar$-lemma of the K\"ahler spaces, see for instance \cite{Cavalcanti:2005hq}, that states that a $\del$-exact and $\delbar$-closed form is $\del\delbar$-exact, we find  that $\del\delta C^{1,3}$ actually must be $\del\delbar$-exact. And therefore  $\delta C^{1,3}=\delbar\Lambda^{1,2}$. Hence, it can be reabsorbed by a gauge transformation $\delta C_{\it 4}\rightarrow \delta C_{\it 4}-\d(\Lambda^{1,2}+\text{c.c.})$.

Now we use (\ref{realinfBI}) to calculate the  the anti-holomorphic variation
\bseq\label{defc}
\begin{align}
\delbar\, \frac{\delta C_{\it 4}}{\delta \hat{\bar z}^{\bar\imath}}&=\,\frac{\delta F^{2,3}}{\delta \hat{\bar z}^{\bar\imath}}\label{defc1}\ ,\\
\del\, \frac{\delta C_{\it 4}}{\delta \hat{\bar z}^{\bar\imath}}&=\,\frac{\delta F^{3,2}}{\delta \hat{\bar z}^{\bar\imath}}+\ell_s^4\iota_{\bar\imath}\delta^{\it 6}_{\hat y}\label{defc2}\ .
\end{align}
\eseq
On the other hand, we have to preserve the bulk condition  $F_{\it 5}=*\d e^{-4A_{\rm E}}$, which translates  into the following expression for the RR five-form
\be\label{holF5}
F^{2,3}=\frac{i}{2}\,\delbar e^{-4A_{\rm E}}\wedge J\wedge J\ .
\ee
Now, by combining (\ref{defc1}) and (\ref{holF5}) we obtain that
\be\label{defRRpot}
\frac{\delta C_{\it 4}}{\delta \hat{\bar z}^{\bar\imath}}=\frac{i}{2}\, \frac{\delta e^{-4A_{\rm E}}}{\delta  \hat{\bar z}^{\bar\imath} }\, J\wedge J+\omega_{\it 4}
\ee
where $\omega_{\it 4}$ is some $(2,2)$ form. Clearly $\omega_{\it 4}$ must be $\delbar$-closed as follows from (\ref{defc1}). Moreover it must be  $\delbar$-exact. To see that we notice that
 freedom in the choice of $\omega_{\it 4}$ is almost completely removed by imposing the compatibility with (\ref{defc2})
\be\label{omegaeq}
\del\omega_{\it 4}=2\frac{\delta F^{3,2}}{\delta \hat{\bar z}^{\bar\imath}}+\ell_s^4\iota_{\bar\imath}\delta^{\it 6}_{\hat y}\ .
\ee
Now, this is the $(3,2)$ form and it is obviously $\del$-closed. That means $\del\omega_{\it 4}$ is also $\del$-exact, simply because the $(3,2)$ cohomology group is trivial for the GKP backgrounds. On the other hand, from (\ref{holF5}) it follows that $\d F_{\it 5}=2\delbar F^{3,2}$ and then  (\ref{omegaeq}) is also $\delbar$-closed. Then we again use the $\del\delbar$-lemma of the K\"ahler spaces to conclude that $\partial \omega_4$ is in fact $\del\delbar$-exact. Hence, $\omega_{\it 4}$ is the $\delbar$-exact form that solves (\ref{omegaeq}) plus some $\d$ closed $(2,2)$ piece.
This closed $(2,2)$ piece is not fixed by the BI and the equations of motion and is related to the gauge freedom in the definition of $C_{\it 4}$. Clearly we require this part to be independent
of the position of the D3-brane $\hat y$.

Now it is time to use (\ref{defRRpot}) to calculate the variation of $\log W_{\rm np}$. Since the $\hat y$ dependent part of $\omega_{\it 4}$ is $\delbar$-exact it does not give any contribution.
Combining together  (\ref{imE3action}) and (\ref{step1}) we get
\be
\frac{\delta \log W_{\rm np}}{\delta \hat{\bar z}^{\bar\imath}}=0\ ,
\ee
which is equivalent to the holomorphy condition (\ref{holsup}). This concludes our derivation of the non-perturbative superpotential (\ref{effsupred}).

\end{appendix}





\begin{thebibliography}{99}

\bibitem{KKLMMT}
  S.~Kachru, R.~Kallosh, A.~D.~Linde and S.~P.~Trivedi,
  ``De Sitter vacua in string theory,''
  Phys.\ Rev.\  D {\bf 68}, 046005 (2003)
  [arXiv:hep-th/0301240].
  S.~Kachru, R.~Kallosh, A.~D.~Linde, J.~M.~Maldacena, L.~P.~McAllister and S.~P.~Trivedi,
  ``Towards inflation in string theory,''
  JCAP {\bf 0310}, 013 (2003)
  [arXiv:hep-th/0308055].






\bibitem{KW}
  I.~R.~Klebanov and E.~Witten,
 ``Superconformal field theory on threebranes at a Calabi-Yau  singularity,''
  Nucl.\ Phys.\  B {\bf 536}, 199 (1998)
  [arXiv:hep-th/9807080].


\bibitem{ks}
  I.~R.~Klebanov and M.~J.~Strassler,
  ``Supergravity and a confining gauge theory: Duality cascades and
  chiSB-resolution of naked singularities,''
  JHEP {\bf 0008}, 052 (2000)
  [arXiv:hep-th/0007191].

\bibitem{Baumann:2007ah}
  D.~Baumann, A.~Dymarsky, I.~R.~Klebanov and L.~McAllister,
  ``Towards an Explicit Model of D-brane Inflation,''
  JCAP {\bf 0801}, 024 (2008)
  [arXiv:0706.0360 [hep-th]].
  D.~Baumann, A.~Dymarsky, I.~R.~Klebanov, L.~McAllister and P.~J.~Steinhardt,
  ``A Delicate Universe,''
  Phys.\ Rev.\ Lett.\  {\bf 99}, 141601 (2007)
  [arXiv:0705.3837 [hep-th]].

\bibitem{cecotti09}
  S.~Cecotti, M.~C.~N.~Cheng, J.~J.~Heckman and C.~Vafa,
  ``Yukawa Couplings in F-theory and Non-Commutative Geometry,''
  arXiv:0910.0477 [hep-th].

\bibitem{npYuk}
  F.~Marchesano and L.~Martucci,
  ``Non-perturbative effects on seven-brane Yukawa couplings,''
  Phys.\ Rev.\ Lett.\  {\bf 104}, 231601 (2010)
  [arXiv:0910.5496 [hep-th]].




\bibitem{gkp}
  S.~B.~Giddings, S.~Kachru and J.~Polchinski,
  ``Hierarchies from fluxes in string compactifications,''
  Phys.\ Rev.\  D {\bf 66} (2002) 106006
  [arXiv:hep-th/0105097].


\bibitem{tentofour}
  P.~Koerber and L.~Martucci,
  ``From ten to four and back again: how to generalize the geometry,''
  JHEP {\bf 0708} (2007) 059
  [arXiv:0707.1038 [hep-th]].


\bibitem{bau10}
  D.~Baumann, A.~Dymarsky, S.~Kachru, I.~R.~Klebanov and L.~McAllister,
  JHEP {\bf 1006}, 072 (2010)
  [arXiv:1001.5028 [hep-th]].




\bibitem{mn}
  J.~M.~Maldacena and C.~Nunez,
  ``Towards the large N limit of pure N = 1 super Yang Mills,''
  Phys.\ Rev.\ Lett.\  {\bf 86} (2001) 588
  [arXiv:hep-th/0008001].

 \bibitem{vafa}
  C.~Vafa,
  ``Superstrings and topological strings at large N,''
  J.\ Math.\ Phys.\  {\bf 42} (2001) 2798
  [arXiv:hep-th/0008142].

\bibitem{vafa2}
  F.~Cachazo, K.~A.~Intriligator and C.~Vafa,
  ``A large N duality via a geometric transition,''
  Nucl.\ Phys.\  B {\bf 603} (2001) 3
  [arXiv:hep-th/0103067].



 \bibitem{martellimalda}
  J.~Maldacena and D.~Martelli,
  ``The unwarped, resolved, deformed conifold: fivebranes and the baryonic
  branch of the Klebanov-Strassler theory,''
  JHEP {\bf 1001} (2010) 104
  [arXiv:0906.0591 [hep-th]].



 \bibitem{ganor96}
  O.~J.~Ganor,
  ``A note on zeroes of superpotentials in F-theory,''
  Nucl.\ Phys.\  B {\bf 499} (1997) 55
  [arXiv:hep-th/9612077].


\bibitem{baumann0607}
  D.~Baumann, A.~Dymarsky, I.~R.~Klebanov, J.~M.~Maldacena, L.~P.~McAllister and A.~Murugan,
  ``On D3-brane potentials in compactifications with fluxes and wrapped
  D-branes,''
  JHEP {\bf 0611} (2006) 031
  [arXiv:hep-th/0607050].





\bibitem{GranaReview}
  M.~Grana,
  ``Flux compactifications in string theory: A comprehensive review,''
  Phys.\ Rept.\  {\bf 423}, 91 (2006)
  [arXiv:hep-th/0509003].




 \bibitem{hitchin}
  N.~Hitchin,
  ``Generalized Calabi-Yau manifolds,''
  Quart.\ J.\ Math.\ Oxford Ser.\  {\bf 54} (2003) 281
  [arXiv:math/0209099].

 \bibitem{gualtieri}
  M.~Gualtieri,
  ``Generalized complex geometry,''
  arXiv:math/0401221.

\bibitem{grana1}
  M.~Grana, J.~Louis, D.~Waldram,
  ``Hitchin functionals in N=2 supergravity,''
  JHEP {\bf 0601}, 008 (2006).
  [hep-th/0505264].


\bibitem{effsugra}
  L.~Martucci,
  ``On moduli and effective theory of N=1 warped flux compactifications,''
  JHEP {\bf 0905} (2009) 027
  [arXiv:0902.4031 [hep-th]].


\bibitem{bw}
  J.~Wess and J.~Bagger,
  ``Supersymmetry and supergravity,''
{\it  Princeton, USA: Univ. Pr. (1992) 259 p}


\bibitem{Veneziano:1982ah}
  G.~Veneziano and S.~Yankielowicz,
  ``An Effective Lagrangian For The Pure N=1 Supersymmetric Yang-Mills
  Theory,''
  Phys.\ Lett.\  B {\bf 113}, 231 (1982).


\bibitem{Lust:2005cu}
  D.~L\"ust, S.~Reffert, W.~Schulgin and P.~K.~Tripathy,
  ``Fermion Zero Modes in the Presence of Fluxes and a Non-perturbative
  Superpotential,''
  JHEP {\bf 0608}, 071 (2006)
  [arXiv:hep-th/0509082].

\bibitem{drsw}
  M.~Dine, R.~Rohm, N.~Seiberg and E.~Witten,
  ``Gluino Condensation In Superstring Models,''
  Phys.\ Lett.\  B {\bf 156} (1985) 55.


\bibitem{hor96}
  P.~${\rm Ho\check{r}ava}$,
  ``Gluino condensation in strongly coupled heterotic string theory,''
  Phys.\ Rev.\  D {\bf 54} (1996) 7561
  [arXiv:hep-th/9608019].

\bibitem{Frey}
  A.~R.~Frey, M.~Lippert,
  ``AdS strings with torsion: Non-complex heterotic compactifications,''
  Phys.\ Rev.\  {\bf D72}, 126001 (2005).
  [hep-th/0507202].

\bibitem{HW}
  P.~${\rm Ho\check{r}ava}$ and E.~Witten,
  ``Heterotic and type I string dynamics from eleven dimensions,''
  Nucl.\ Phys.\  B {\bf 460} (1996) 506
  [arXiv:hep-th/9510209].
  P.~${\rm Ho\check{r}ava}$ and E.~Witten,
  ``Eleven-Dimensional Supergravity on a Manifold with Boundary,''
  Nucl.\ Phys.\  B {\bf 475} (1996) 94
  [arXiv:hep-th/9603142].


\bibitem{nilles}
  K.~A.~Meissner, H.~P.~Nilles, M.~Olechowski,
  ``Supersymmetry breakdown at distant branes: The super Higgs mechanism,''
  Nucl.\ Phys.\  {\bf B561 } (1999)  30-42.
  [hep-th/9905139].

\bibitem{candelas}
  P.~Candelas and X.~C.~de la Ossa,
  ``Comments on Conifolds,''
  Nucl.\ Phys.\  B {\bf 342} (1990) 246.

\bibitem{lucal}
  L.~Martucci and P.~Smyth,
  ``Supersymmetric D-branes and calibrations on general N = 1 backgrounds,''
  JHEP {\bf 0511} (2005) 048
  [arXiv:hep-th/0507099].
  
\bibitem{thesis}
 A.~Dymarsky, ``Warped Throat Solutions in String Theory
and Their Cosmological Applications,'' Ph.D. thesis, Princeton University 2007

\bibitem{buttietal}
  A.~Butti, M.~Grana, R.~Minasian, M.~Petrini and A.~Zaffaroni,
  ``The baryonic branch of Klebanov-Strassler solution: A supersymmetric
  family of SU(3) structure backgrounds,''
  JHEP {\bf 0503}, 069 (2005)
  [arXiv:hep-th/0412187].


 \bibitem{gvw}
  S.~Gukov, C.~Vafa, E.~Witten,
  ``CFT's from Calabi-Yau four folds,''
  Nucl.\ Phys.\  {\bf B584}, 69-108 (2000).
  [hep-th/9906070].



\bibitem{MP}
  D.~R.~Morrison and M.~R.~Plesser,
  ``Non-spherical horizons. I,''
  Adv.\ Theor.\ Math.\ Phys.\  {\bf 3}, 1 (1999)
  [arXiv:hep-th/9810201].

\bibitem{DKS}
  A.~Dymarsky, I.~R.~Klebanov and N.~Seiberg,
  ``On the moduli space of the cascading SU(M+p) x SU(p) gauge theory,''
  JHEP {\bf 0601}, 155 (2006)
  [arXiv:hep-th/0511254].



\bibitem{atyah}
  M.~Atiyah, J.~M.~Maldacena and C.~Vafa,
  ``An M-theory flop as a large N duality,''
  J.\ Math.\ Phys.\  {\bf 42} (2001) 3209
  [arXiv:hep-th/0011256].

\bibitem{paulreview}
  P.~Koerber,
  ``Lectures on Generalized Complex Geometry for Physicists,''
  arXiv:1006.1536 [hep-th].

\bibitem{gmpt}
  M.~Grana, R.~Minasian, M.~Petrini and A.~Tomasiello,
  ``Generalized structures of N=1 vacua,''
  JHEP {\bf 0511} (2005) 020
  [arXiv:hep-th/0505212].


\bibitem{Held:2010az}
  J.~Held, D.~L\"ust, F.~Marchesano and L.~Martucci,
  ``DWSB in heterotic flux compactifications,'' 
  JHEP {\bf 1006}, 090 (2010)
  [arXiv:1004.0867 [hep-th]].

\bibitem{pauldimi}
  P.~Koerber and D.~Tsimpis,
  ``Supersymmetric sources, integrability and generalized-structure
  compactifications,''
  JHEP {\bf 0708} (2007) 082
  [arXiv:0706.1244 [hep-th]].

\bibitem{cacha01}
  F.~Cachazo, B.~Fiol, K.~A.~Intriligator, S.~Katz and C.~Vafa,
  ``A geometric unification of dualities,''
  Nucl.\ Phys.\  B {\bf 628} (2002) 3
  [arXiv:hep-th/0110028].

\bibitem{franco10}
  S.~Franco and G.~Torroba,
  ``Gauge theories from D7-branes over vanishing 4-cycles,''
  arXiv:1010.4029 [hep-th].



  \bibitem{lucasup}
  L.~Martucci,
  ``D-branes on general N = 1 backgrounds: Superpotentials and D-terms,''
  JHEP {\bf 0606} (2006) 033
  [arXiv:hep-th/0602129].

\bibitem{GH}
P.~Griffiths and J.~Harris, ``Principles of algebraic geometry", Wiley-Interscience Publication, 1994




\bibitem{Dbraneferm}
D.~Marolf, L.~Martucci and P.~J.~Silva,
  ``Fermions, T-duality and effective actions for D-branes in bosonic
  backgrounds,''
  JHEP {\bf 0304} (2003) 051
  [arXiv:hep-th/0303209];
D.~Marolf, L.~Martucci and P.~J.~Silva,
  ``Actions and fermionic symmetries for D-branes in bosonic backgrounds,''
  JHEP {\bf 0307} (2003) 019
  [arXiv:hep-th/0306066];
  L.~Martucci, J.~Rosseel, D.~Van den Bleeken and A.~Van Proeyen,
  ``Dirac actions for D-branes on backgrounds with fluxes,''
  Class.\ Quant.\ Grav.\  {\bf 22} (2005) 2745
  [arXiv:hep-th/0504041].



\bibitem{camara}
  P.~G.~Camara, L.~E.~Ibanez, A.~M.~Uranga,
  ``Flux-induced SUSY-breaking soft terms on D7-D3 brane systems,''
  Nucl.\ Phys.\  {\bf B708 } (2005)  268-316.
  [hep-th/0408036].




\bibitem{dwsb}
  D.~L\"ust, F.~Marchesano, L.~Martucci and D.~Tsimpis,
  ``Generalized non-supersymmetric flux vacua,''
  JHEP {\bf 0811} (2008) 021
  [arXiv:0807.4540 [hep-th]].

\bibitem{heid10}
  B.~Heidenreich, L.~McAllister and G.~Torroba,
  ``Dynamic SU(2) Structure from Seven-branes,''
  arXiv:1011.3510 [hep-th].


\bibitem{Butti:2007aq}
  A.~Butti, D.~Forcella, L.~Martucci, R.~Minasian, M.~Petrini and A.~Zaffaroni,
  ``On the geometry and the moduli space of beta-deformed quiver gauge
  theories,'' 
  JHEP {\bf 0807}, 053 (2008)
  [arXiv:0712.1215 [hep-th]].

\bibitem{Lunin:2005jy}
  O.~Lunin and J.~M.~Maldacena,
  ``Deforming field theories with U(1) x U(1) global symmetry and their
  gravity duals,''
  JHEP {\bf 0505}, 033 (2005)
  [arXiv:hep-th/0502086].


\bibitem{Tomasiello:2007zq}
  A.~Tomasiello,
  ``Reformulating Supersymmetry with a Generalized Dolbeault Operator,''
  JHEP {\bf 0802}, 010 (2008)
  [arXiv:0704.2613 [hep-th]].


\bibitem{LARGE}
  V.~Balasubramanian, P.~Berglund, J.~P.~Conlon and F.~Quevedo,
  ``Systematics of Moduli Stabilisation in Calabi-Yau Flux Compactifications,''
  JHEP {\bf 0503} (2005) 007
  [arXiv:hep-th/0502058].

\bibitem{wittensup}
  E.~Witten,
  ``Non-Perturbative Superpotentials In String Theory,''
  Nucl.\ Phys.\  B {\bf 474} (1996) 343
  [arXiv:hep-th/9604030].

\bibitem{beckerflux}
  K.~Becker and M.~Becker,
  ``M-Theory on Eight-Manifolds,''
  Nucl.\ Phys.\  B {\bf 477} (1996) 155
  [arXiv:hep-th/9605053];
  K.~Becker and M.~Becker,
  ``Supersymmetry breaking, M-theory and fluxes,''
  JHEP {\bf 0107} (2001) 038
  [arXiv:hep-th/0107044].


\bibitem{Giddings:2005ff}
  S.~B.~Giddings and A.~Maharana,
  ``Dynamics of warped compactifications and the shape of the warped
  landscape,''
  Phys.\ Rev.\  D {\bf 73}, 126003 (2006)
  [arXiv:hep-th/0507158].


\bibitem{Cavalcanti:2005hq}
  G.~R.~Cavalcanti,
  ``New aspects of the ddc-lemma,''
  arXiv:math/0501406.

\end{thebibliography}
\end{document}